\documentclass[journal=nalefd,manuscript=article]{achemso}

\usepackage[T1]{fontenc} 
\usepackage{placeins}
\usepackage[lofdepth,lotdepth]{subfig}
\setkeys{acs}{articletitle=true}

\author{Davide Campi}
\affiliation[1]{Theory and Simulation of Materials (THEOS), and National Centre for Computational Design and Discovery of Novel Materials (MARVEL), \'Ecole Polytechnique F\'ed\'erale de Lausanne, CH-1015 Lausanne, Switzerland}
\affiliation{Dipartimento di Scienza dei Materiali, University of Milano-Bicocca, Via R.Cozzi 55, Milano, Italy}
\email{davide.campi@unimib.it}
\author{Nicolas Mounet}
\affiliation{Theory and Simulation of Materials (THEOS), and National Centre for Computational Design and Discovery of Novel Materials (MARVEL), \'Ecole Polytechnique F\'ed\'erale de Lausanne, CH-1015 Lausanne, Switzerland}
\author{Marco Gibertini}
\affiliation{Theory and Simulation of Materials (THEOS), and National Centre for Computational Design and Discovery of Novel Materials (MARVEL), \'Ecole Polytechnique F\'ed\'erale de Lausanne, CH-1015 Lausanne, Switzerland}
\affiliation{Dipartimento di Scienze Fisiche, Informatiche e Matematiche, University of Modena and Reggio Emilia, I-41125 Modena, Italy}
\affiliation{Centro S3, Istituto di Nanoscienze - CNR, I-41125, Modena, Italy }
\author{Giovanni Pizzi}
\affiliation{Theory and Simulation of Materials (THEOS), and National Centre for Computational Design and Discovery of Novel Materials (MARVEL), \'Ecole Polytechnique F\'ed\'erale de Lausanne, CH-1015 Lausanne, Switzerland}
\affiliation{Laboratory for Materials Simulations (LMS), Paul Scherrer Institut, CH-5232 Villigen PSI, Switzerland}
\author{Nicola Marzari}
\affiliation{Theory and Simulation of Materials (THEOS), and National Centre for Computational Design and Discovery of Novel Materials (MARVEL), \'Ecole Polytechnique F\'ed\'erale de Lausanne, CH-1015 Lausanne, Switzerland}
\affiliation{Laboratory for Materials Simulations (LMS), Paul Scherrer Institut, CH-5232 Villigen PSI, Switzerland}
\email{nicola.marzari@epfl.ch}

\title{Novel materials in the Materials Cloud 2D database}

\begin{document}

\begin{abstract}

Two-dimensional (2D) materials are among the most promising candidates for beyond-silicon electronic, optoelectronic and quantum computing
applications. Recently, their recognized importance sparked a push to discover and characterize novel 2D materials. Within a few years, the number of experimentally exfoliated or synthesized 2D materials went from a couple of dozens to more than a hundred, with the number of theoretically predicted compounds reaching a few thousands. In 2018 we first contributed to this effort with the identification of 1825 compounds that are
either easily (1036) or potentially (789) exfoliable from experimentally known 3D compounds. Here, we report on a major expansion of this 2D portfolio thanks to the extension of the screening protocol to an additional experimental database (MPDS) as well as to the updated versions of the two databases (ICSD and COD) used in our previous work. This expansion has led to the discovery of additional 1252 unique monolayers, bringing the total to 3077 compounds and, notably, almost doubling the number of easily exfoliable materials (2004). 
Moreover, we optimized the structural properties of all these monolayers and explored their electronic structure with a particular emphasis on those rare large-bandgap 2D materials that could be precious to isolate 2D field effect transistors channels.
Finally, for each material containing up to 6 atoms per unit cell, we   
identified the best candidates to form commensurate heterostructures, balancing requirements on the supercells size and minimal strain.

\end{abstract}

\section{Introduction}

Two-dimensional (2D) materials represent a vast and broadly unexplored region of the materials space. Thanks to their extreme thinness 
they are regarded as ideal platforms for electronic and optoelectronic applications in the beyond-silicon era \cite{Radisavljevic:2011,Wang2012,Liu:2014, Roy2014, Chhowalla2016, Klinkert2020} as well as precious candidates for electro- and photo-catalysis\cite{He2012,Sun2012} and for the realization of exotic states of matter\cite{Qian2016,Marrazzo2018,Campi2021}. Moreover, their properties proved to be much easier to tune with strain, electric fields, and doping with respect to their bulk counterparts, and they can be combined in virtually endless van-der-Waals (vdW) heterostructures~\cite{geim2013van} to engineer novel functionalities.   
Although until few years ago only a few dozens of 2D materials had been actively studied, in recent years major progress has taken place in the experimental synthesis or exfoliation of a variety of more than a hundred 2D materials~\cite{Backes2019,Huang2020}. Even more impressively, the number of 2D materials theoretically predicted using high-throughput computational methods\cite{Gould2016}  has grown from less than two hundred\cite{Nieminen,rasmussen2015computational} to the range of the thousands\cite{Choudhary2017,Ashton2017,Cheon2017,Mounet2018,Haastrup2018,Zhou2019}.    
In our first contribution to this effort\cite{Mounet2018} we screened the Inorganic Crystal Structure Database~\cite{ICSD} (ICSD) and the Crystallography Open Database~\cite{COD} (COD) for layered materials, identifying a total of 1825 2D materials that, on the basis of their computed binding energies, could be exfoliated from their layered parent structures with mechanical\cite{Novoselov2004} or liquid-phase\cite{Coleman2011} exfoliation methods. 
In this work we follow a protocol very similar to the one used in Ref.\cite{Mounet2018} (starting from a geometric 
and bonding criteria to identify layered materials, followed by the calculation of binding energies using first-principles vdW density-functional theory (DFT) simulations), but we add to the sources a third database, the Pauling File~\cite{Villars:1998} (MPDS), and we repeat the screening on ICSD and COD using their most up-to-date versions, as well as allowing the inclusion of larger structures and using slightly less stringent thresholds on the geometrical selection. The latter condition results in a more inclusive selection, at the price of a slightly larger rate of false positives that are later ruled out by DFT calculations. This extended screening allows us to identify 1252 additional 2D materials bringing the total to 3077 and, notably, doubling to 2004 the number of compounds that should be most easily exfoliable. Furthermore, for each of these 3077 monolayers  we optimize the cell and the internal geometry treating it as an isolated 2D system and we compute its electronic band structure. On the basis of their optimized geometry we suggest, for each material up to 6 atoms per unit cell, ideal candidates to build simple, commensurate vertical heterostructures or lattice-matched lateral heterostructures with minimal strain. Finally, we study a handful of materials with exceptionally large bandgaps that could serve as insulating layers for 2D Field Effect Transistor (FET) channels with superior performance with respect to the widely used BN\cite{Knobloch2021}. The full reproducibility of the study is ensured by the AiiDA~\cite{AiiDA,Huber2020} materials' informatics infrastructure, which keeps track of the provenance of each calculation; therefore the results are openly available together with their entire provenance through the Materials Cloud.\cite{Talirz2020,Campi2022MC}.

\section{Results and discussion}
\subsection{Identification of layered materials and optimization of bulk compounds}
Following the recipes and tools detailed in Ref.\cite{Mounet2018} we start the computational exfoliation protocol by extracting the bulk 3D crystal structures, in the form of CIF files~\cite{CIF}, from three experimental repositories: ICSD~\cite{ICSD}, COD~\cite{COD} and the Pauling File (MPDS) \cite{Villars:1998}. We exclude structures with partial occupations together with CIF files that do not provide the explicit positions of one or several atoms, cannot be parsed, or are obviously wrong. Theoretically predicted structures are also discarded when signaled. This results in a total of  $147731$ structural entries for ICSD, $279885$ for COD and $355016$ for MPDS. In this work we focus on entries containing at most $6$ different species and $100$ atoms or less in the primitive unit cell, reducing the number of entries to $140586$, $91161$ and $262010$  for  ICSD, COD and MPDS, respectively. The CIF files are extracted and then converted into AiiDA structures using pymatgen~\cite{pymatgen}.
All these 3D structures are separately analyzed to find possible candidates for exfoliation using the same geometrical screening procedure originally described in Ref.\cite{Mounet2018}, building a connection between two atoms when their distance is smaller than the sum of their respective VdW radii at least by  $\Delta$, which is a parameter in the protocol. 
In the current screening we assume slightly larger uncertainties in the VdW radii\cite{alvarez}, thus allowing $\Delta$ to range between $1.0$~\AA\ and $1.5$~\AA\ ($1.1$~\AA\ - $1.5$~\AA\ was used in the original screening).
The geometrical selection identifies thus a total of 8963 layered materials in ICSD, 6794 in COD, and 11530 in MPDS. These selected structures are then processed with the spglib software~\cite{spglib} to find the primitive cells, and filtered for uniqueness (separately for each source) using the pymatgen structure matcher\cite{CMPZ}. Finally a second cutoff on the number of atoms ( $\leq$ 40 atoms/unit cell independently of the number of atomic species) has been applied, leaving respectively 6933, 6283 and 5907 structures for the three databases. These structures coming from the three different databases have been combined and filtered for uniqueness a second time (this time across the different databases) giving a total of 9306 layered candidates, 3689 of which were not included in our previous screening\cite{Mounet2018}. In Ref. \cite{Mounet2018} the structures of the layered materials obtained from the geometrical selection were optimized  using two different non-local vdW-compliant functionals:  the vdW-DF2 functional~\cite{lee2010} with C09 exchange~\cite{cooper2010,hamada2010} (DF2-C09) and the revised Vydrov-Van Voorhis~\cite{vydrov2009,vydrov2010,sabatini2013} (rVV10) functional. Subsequently the binding energy (the difference per unit area between the total energy of optimized 3D bulk structure and the sum of the energies of each isolated substructure of any dimensionality\cite{bjorkman2012}) was computed with both functionals. The two resulting binding energies turned out to be rather similar and very rarely changed the classification of a material. For this reason we abandon here this redundancy and employ only the vdW DF2-c09 functional for the structural optimization and subsequently the calculation of the binding energies for the new 3689 entries.

\subsection{Binding energies}
\begin{figure}[h!]
\includegraphics[width=0.6\textwidth]{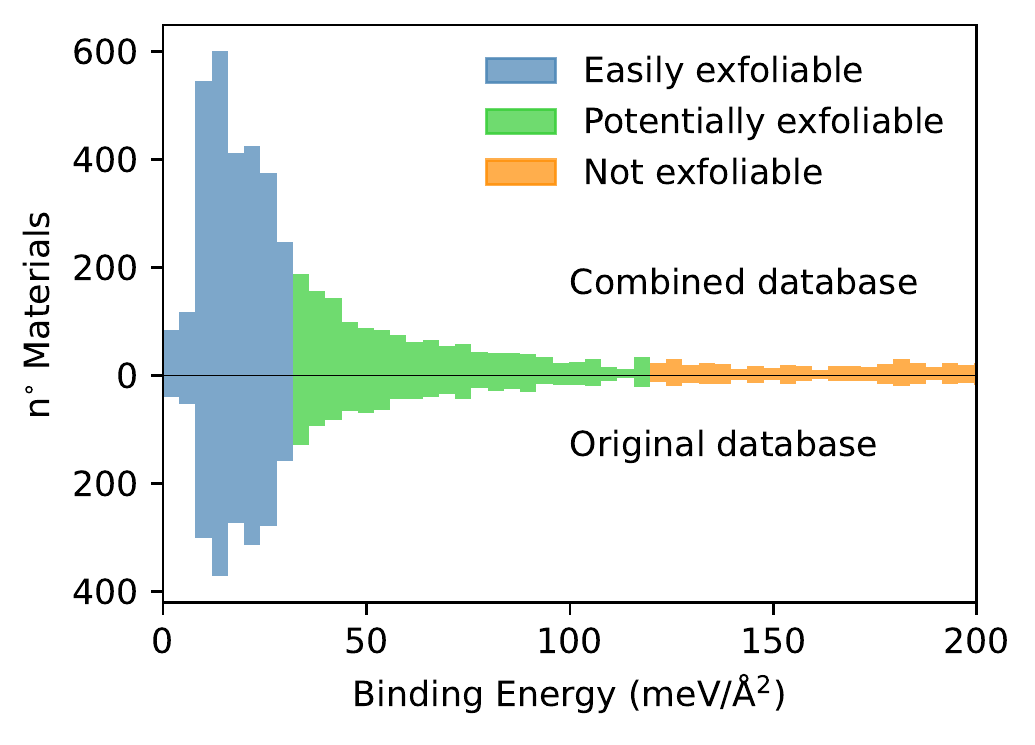}
\caption{Distribution of the 5461 binding energies computed in the combined (MC2D) database compared with the distribution of the binding energies computed in our previous study\cite{Mounet2018}. The color code reflects the classification of easily exfoliable (binding energies $\leq$ 30 meV/\AA$^{2}$) potentially exfoliable (binding energies ranging from 30 to 120 meV/\AA$^{2}$) and non-exfoliable (binding energies larger than 120 meV/\AA$^{2}$).}
\label{fig:BEdist}
\end{figure}

Before calculating the binding energies, all the structures, optimized with vdW DF2-c09, are further screened with the geometrical selection algorithm to assess whether they maintain their layered nature after relaxation. This further selection, together with some unavoidable convergence failures, resulted in a total of $2251$ binding energies successfully computed, to be added to the $3210$ computed in our previous work\cite{Mounet2018}. 
We report in Fig.~\ref{fig:BEdist} the distribution of the binding energies for the new combined database compared with the distribution of the $3210$ binding energies obtained in Ref.\cite{Mounet2018}. The colors reflect the classification of the materials developed in Ref.\cite{Mounet2018} into three classes according to their binding energies: easily exfoliable (with binding energies below 30 meV/\AA$^{2}$, which is close to that of materials routinely exfoliated with standard techniques like graphene or MoS$_2$), potentially exfoliable (binding energies between 30 and 120 meV/\AA$^{2}$) and non-exfoliable (larger than 120 meV/\AA$^{2}$).

Similar to Ref.\cite{Mounet2018}, the majority of the materials exhibit binding energies below 30 meV/\AA$^{2}$.  In the current work the predominance of materials  
with a low binding energy is even more prominent, with a sharper main peak positioned at lower energies and a relatively faster decay of the tail in the potentially exfoliable region. A similar plateau, extending up to 400 meV/\AA$^{2}$, can instead be observed in the non-exfoliable region.

\begin{figure}[h!]
\includegraphics[width=0.6\textwidth]{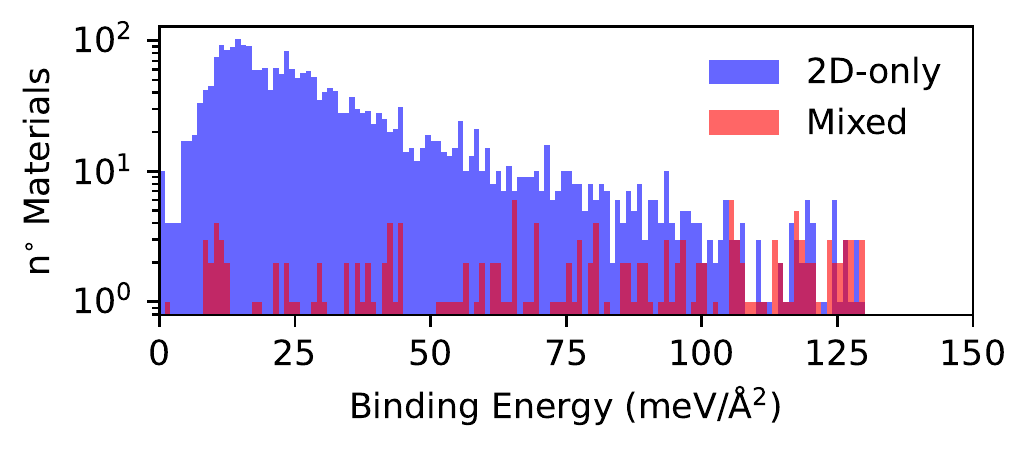}
\caption{Distribution of the binding energies computed in this work classified according to the dimensionality of the bulk parent. In blue are reported the binding energies of materials coming from bulk parents composed only of 2D substructures, while in red are reported materials  deriving from parents with a mixed dimensionality, i.e., including also 1D or 0D structures beside the 2D layers.}
\label{fig:mixed2d}
\end{figure} 

Similarly to our previous results\cite{Mounet2018} we can note that, in percentage,  the region of lower binding energies is largely dominated by materials whose parent 3D structure is composed exclusively of 2D substructures; moving towards higher binding energies materials with mixed dimensionality (typically 0D and 2D) become more common (see Fig.~\ref{fig:mixed2d}), even if with less  prominence  than what previously observed. These distributions can be rationalized considering that charge transfer in layered materials, between intercalated units and 2D layers, can take place giving rise to stronger ionic bonds.

All the 2D substructures obtained from the exofoliation process are filtered for duplicates: it is in fact common that the same monolayer can be obtained from different bulk parents. Each monolayer can therefore be associated with multiple binding energies, according to the parent they have been isolated from. For practical purposes, in the analysis that follows we associate each unique monolayer to the lowest possible binding energy.  Upon removal of duplicates, we end up with $1252$ new unique monolayers, to be added to the $1825$ found in our previous work\cite{Mounet2018}. Remarkably, a large number of these materials fall in the easily exfoliable category, $968$ out of $1252$, representing more than $77\%$ of the materials, a significantly larger percentage than the $57\%$ found in our previous study.  The net result is that we have now doubled the number of easily exfoliable materials, bringing the total to $2004$ in the combined database. These represent an optimal target for future experimental or theoretical studies, driven by the attractive combination of having an established experimental bulk parent, and being easily exfoliable.

\begin{figure}[h!]
\includegraphics[width=0.6\textwidth]{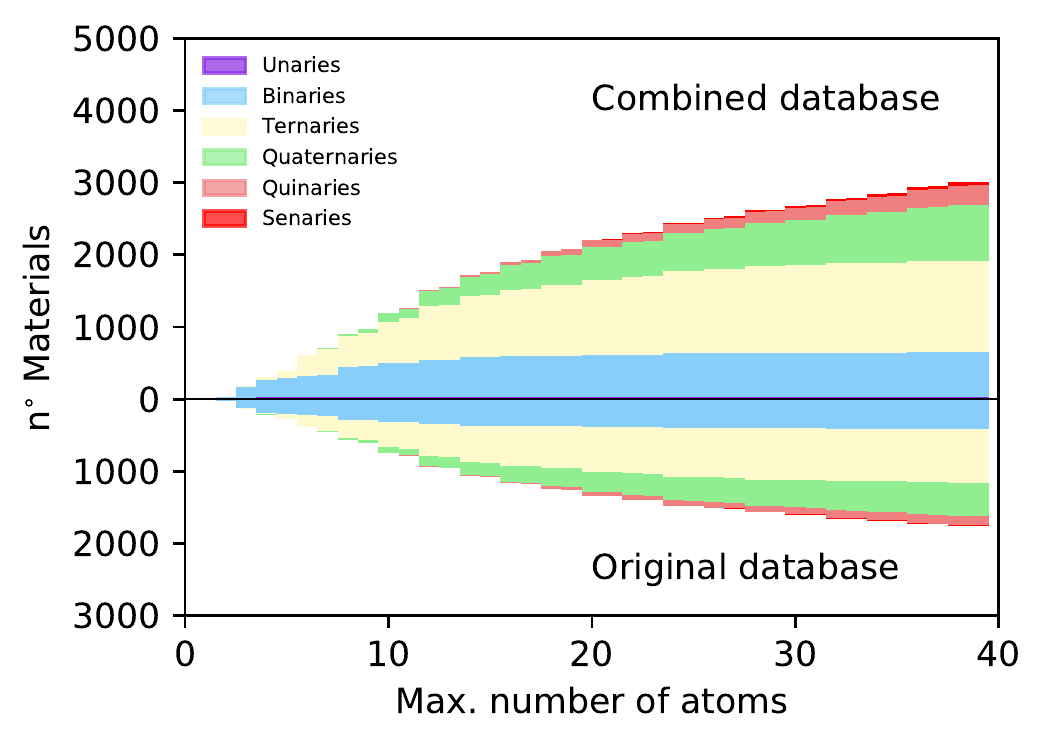}
\caption{Number of structures as a function of the number of atoms in the primitive cell (top: exfoliable 2D structures in the overall database; bottom: exfoliable 2D structures obtained in Ref. \cite{Mounet2018} ).}
\label{fig:Natdist}
\end{figure}

In Fig.~\ref{fig:Natdist} we show how the 2D compounds are distributed in terms of the number of species and number of atoms in the unit cell in the combined database compared with old results. Overall, the distributions are fairly similar, in proportion among the new structures there are less unary and binary structures with small unit cells (up to 6 atoms per unit cell). This is to be expected since the simpler unary and binary monolayers as well as their parent compounds have been heavily studied and therefore had a higher chance to be included in our previous screening. Nevertheless we found 10 new unary compounds. The greater number of quinaries and senaries materials with larger unit cells reflects instead the choice to include in the current screening all the materials up to 40 atoms per unit cell regardless of the number of species, while previously a cutoff of 32 atoms per unit cell had been applied for materials with more than 4 atomic types.   

\begin{figure}[!ht]
\centering
\centering
\subfloat[][]{
\includegraphics[angle=0, width=0.35\textwidth]{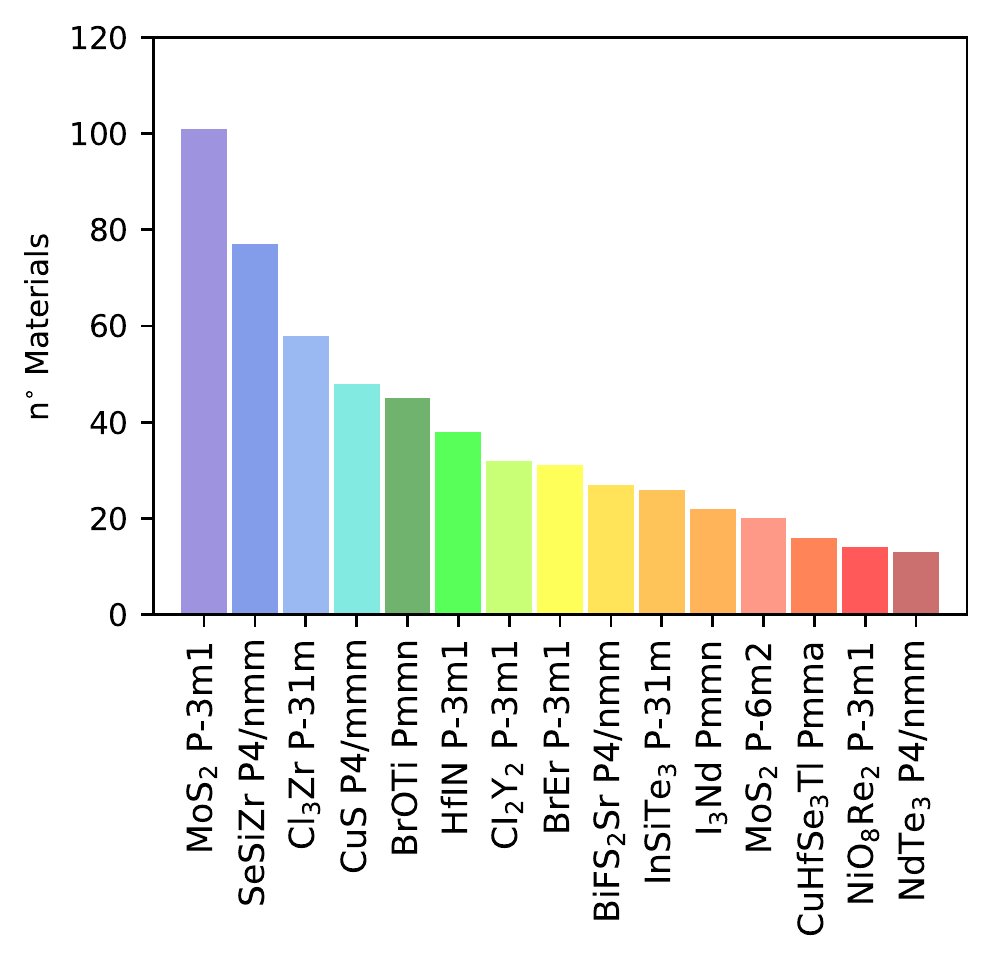}
\label{fig:Prdist}
}
\subfloat[][]{
\includegraphics[angle=0, width=0.65\textwidth]{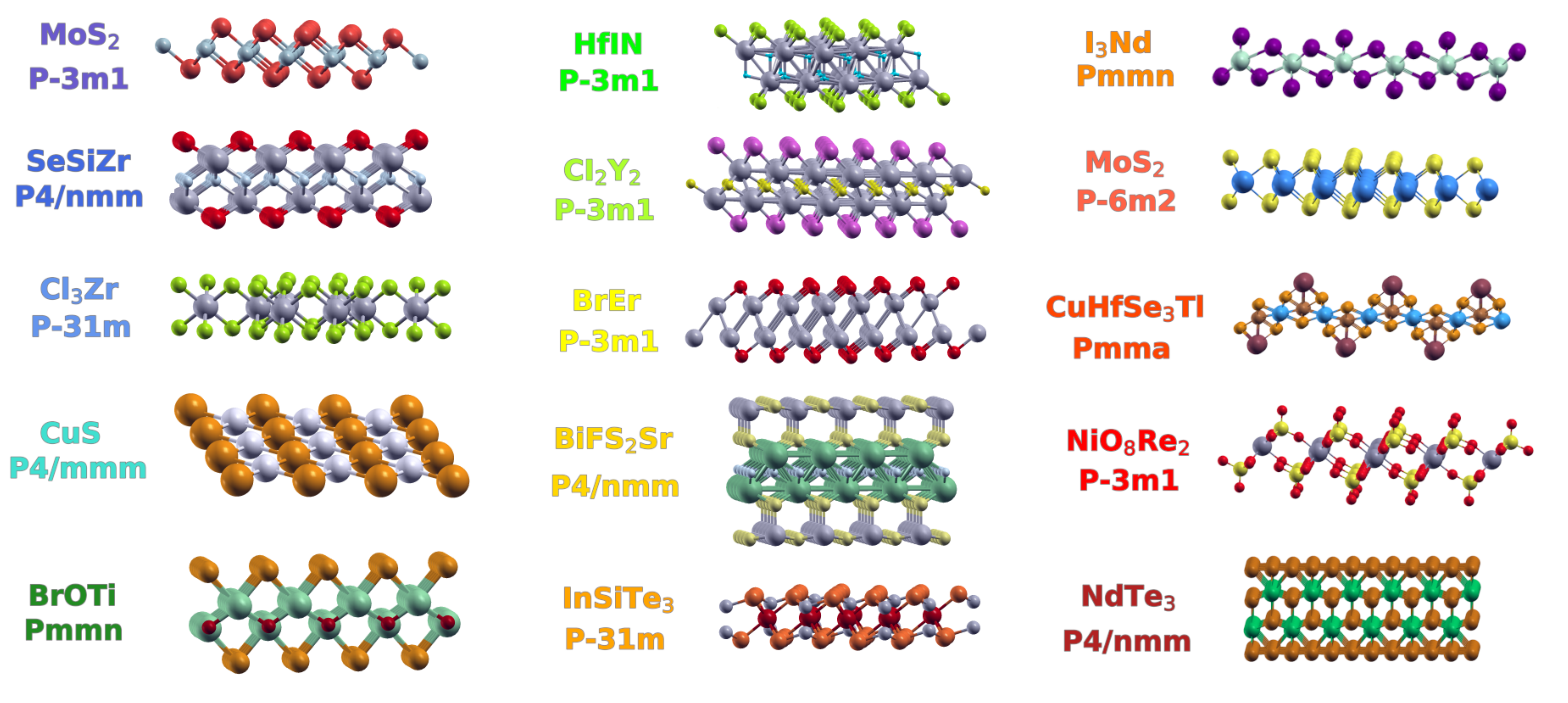}
\label{fig:Prfigs}
}
\caption[]{ \subref{fig:Prdist} Most common 2D structural prototypes:  histogram with the number of structures belonging to the fifteen most common 2D structural prototypes in the database regardless of their classification as easily or potentially exfoliable. \subref{fig:Prfigs} Graphical representation of each prototype, together with the structure-type formula and the spacegroup of the 2D systems.
}
\label{fig:2Dprototypes}
\end{figure} 

In order to provide a more precise overview, we classify the 2D materials  into different prototypes, according to their spacegroups and their structural similarity, considering all the elements indistinguishable. In the combined database we find a total of $1124$
prototypes, effectively doubling the number of prototypes found in our previous screening ($566$). The fifteen most common prototypes, accounting for a total of 568 structures, are reported in Fig.~\ref{fig:2Dprototypes}. The most common structural prototype is the one including transition-metal dichalcogenides and dihalides such as MoS$_2$ and CdI$_2$ in the 1T structure with the hexagonal space group $P{-}3m1$; this counts $102$ similar structures. In second and third position with $77$ and $58$ representatives, we find the rectangular SeSiZr prototype and the hexagonal trihalides, many of which, like CrI$_3$, are currently heavily studied for their outstanding magnetic properties~\cite{McGuire2017}. Surprisingly, 2H-TMDs are only ranked 12$^{\rm th}$ with 20 structures. 
Eighth in order of abundance (with $31$ structures) we find a previously unreported prototype of a binary rare-earth/halide structure represented in Fig.~\ref{fig:2Dprototypes} by BrEr.
 
\begin{figure}[h!]
\includegraphics[width=0.6\textwidth]{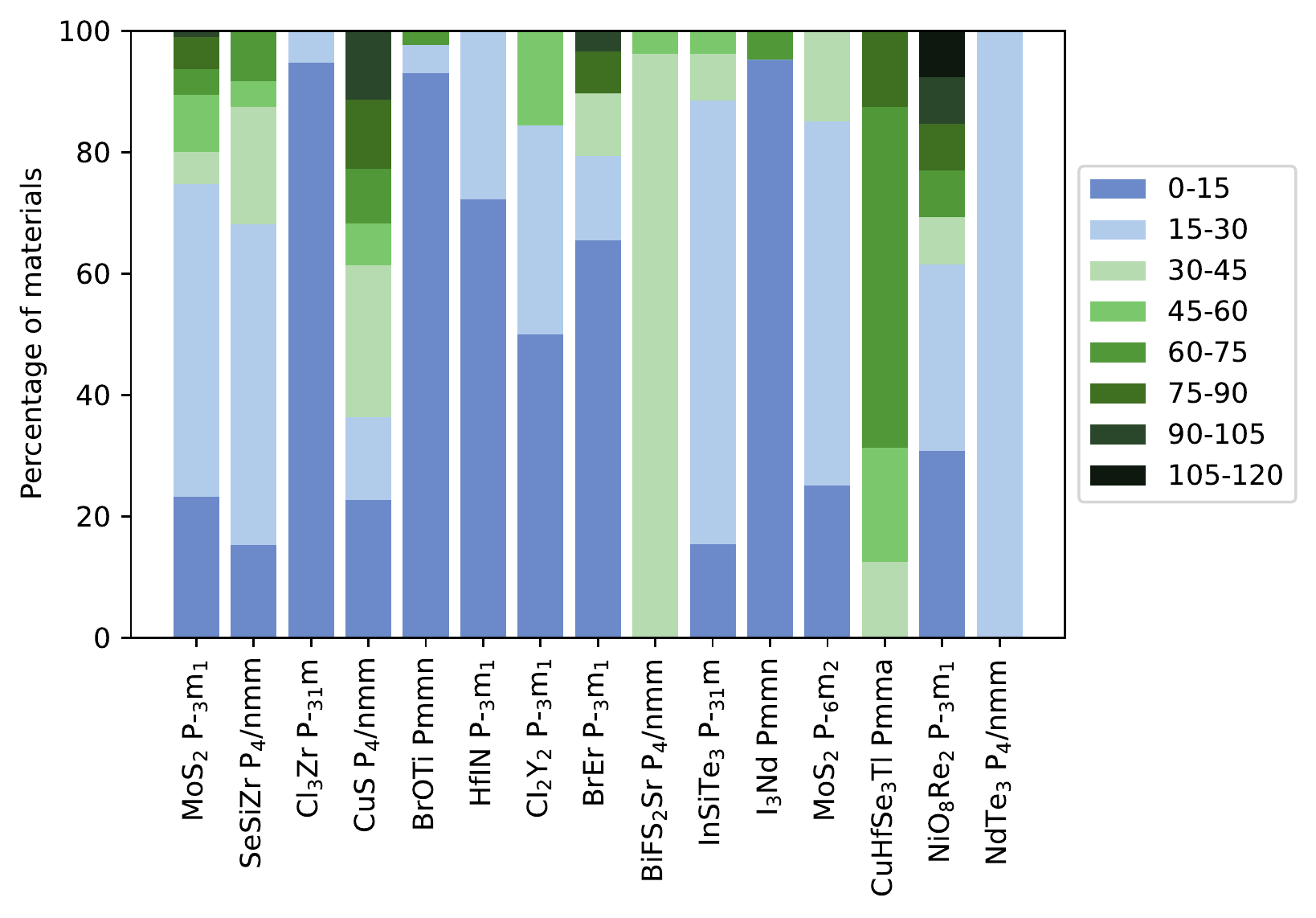}
\caption{Distribution of the binding energies for the materials in the fifteen most common prototype classes. The $0{-}12$0 meV/\AA$^{2}$ range has been equally divided in eight parts each represented by a different color.}
\label{fig:protoBE}
\end{figure} 

To provide further insight, we report in Fig.~\ref{fig:protoBE} the distribution of the binding energies for the fifteen prototype classes enumerated in Fig.~\ref{fig:2Dprototypes}. It can be noted that while some classes, like tri-halides and tri-chalcogenides, present an almost uniform distribution of the binding energies, in many cases the classes span the entire range of binding energies with larger binding energies often associated with heavier atoms in the outermost layers. The relative abundance of high binding energies in the class of square-lattice transition-metal monochalcogenides represented by CuS explains why this prototype does not appear in our previous analysis\cite{Mounet2018}, which was focused exclusively on easily exfoliable materials. The same holds for the prototypes represented by BiFS$_2$Sr, important for its potential applications in spin-FET transistors, as well as the one represented by CuHfSe$_3$Tl that contains exclusively materials in the potentially exfoliable category.

\subsection{Optimization of the 2D monolayers}
\begin{figure}[!ht]
\centering
\centering
\subfloat[][]{
\includegraphics[angle=0, width=0.48\textwidth]{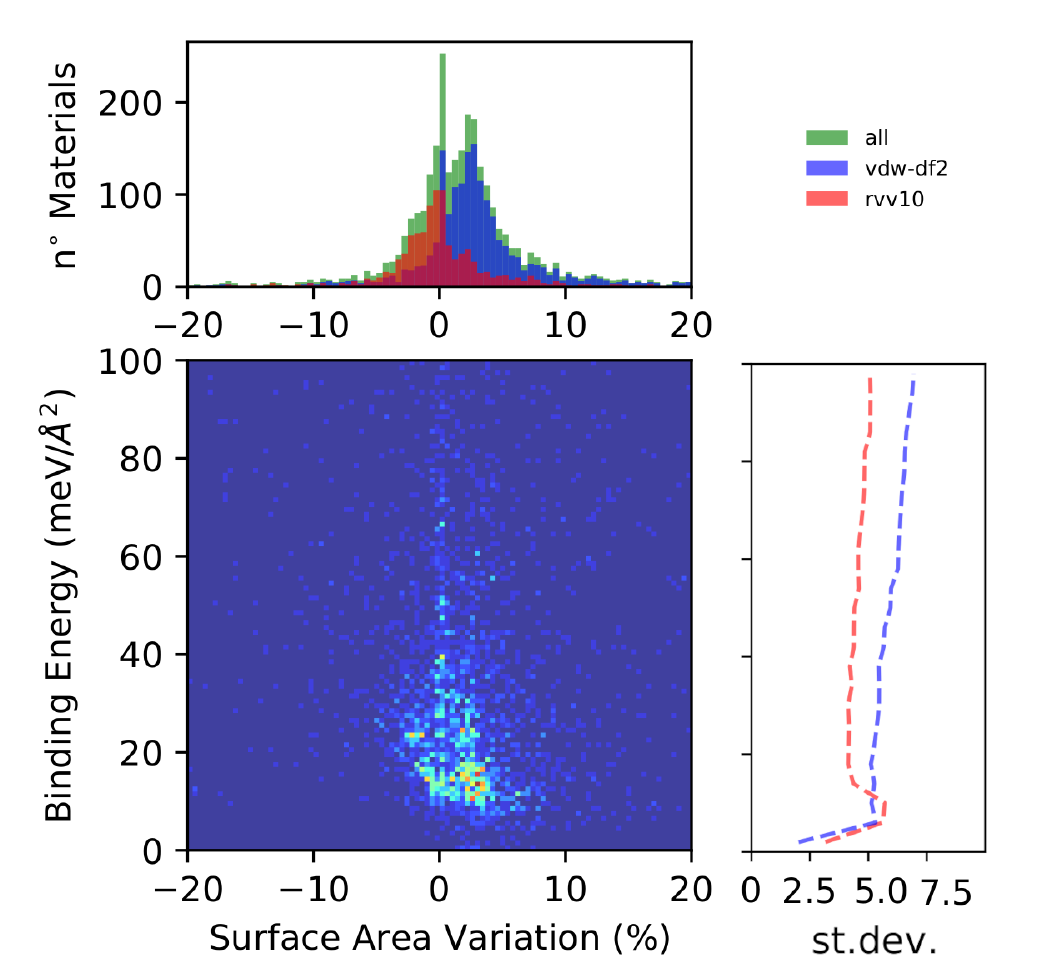}
\label{fig:DShist}
}
\subfloat[][]{
\includegraphics[angle=0, width=0.5\textwidth]{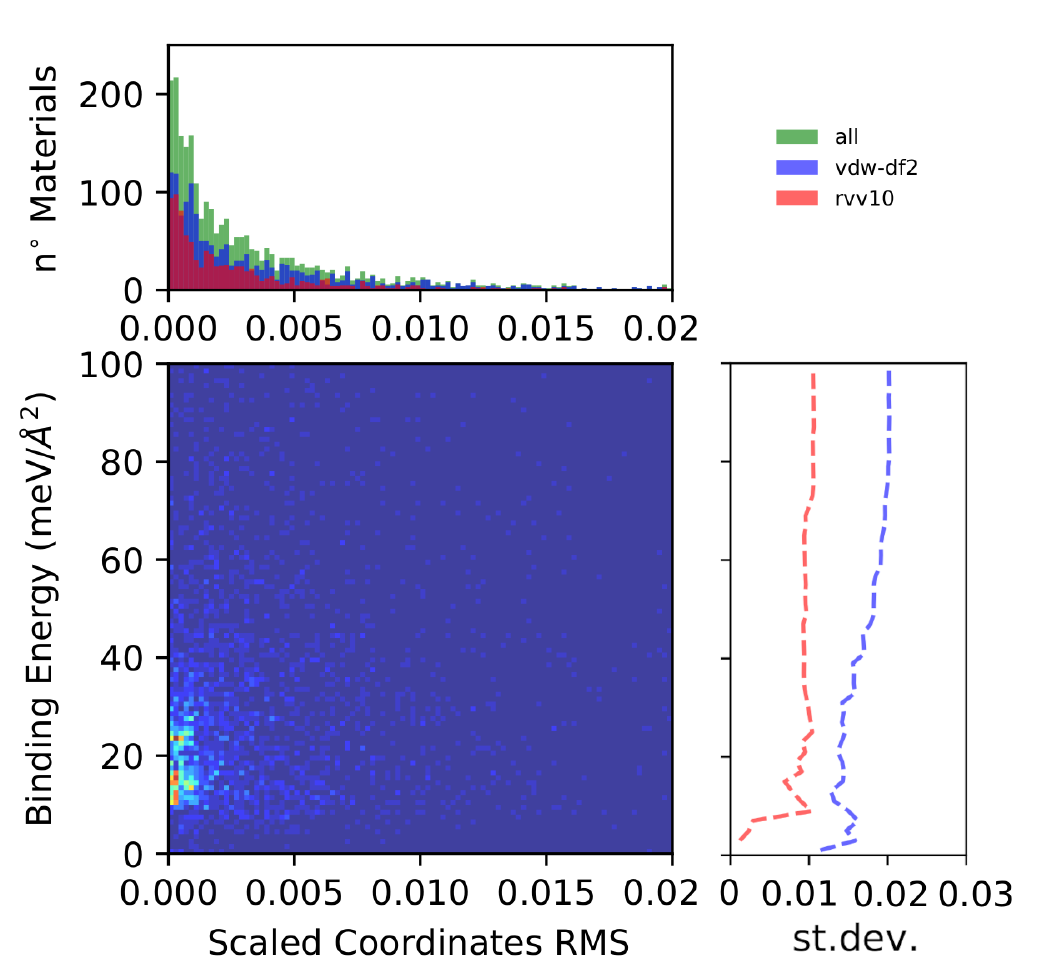}
\label{fig:RMShist}
}
\caption[]{ \subref{fig:DShist} Distribution of the surface area variation during the variable cell relaxation as isolated 2D structures with PBE functional. In green all the structures, in blue the structures whose 3D parent has been relaxed with vdW DF2-c09, in red the ones whose 3D parent has been relaxed with rVV10 in Ref.\cite{Mounet2018}. Below we report a heatmap representing the distribution of the number of structures accordingly to their binding energy and the variation of their surface area during the optimization as isolated 2D materials. Beside we report the normalized standard deviation of the surface variation distribution as a function of the binding energy , a slow but steady increase can be observed.\label{fig:dsbe}
\subref{fig:RMShist} Distribution, heatmap and evolution of the normalized standard deviation for the root mean square (RMS) of the scaled positions during the relaxation.  
}
\label{fig:relax}
\end{figure}

All the 3077 monolayers composing the combined database have been optimized as isolated 2D structures performing a variable cell relaxation in open-boundary conditions\cite{Sohier2015} and using the PBE\cite{PBE} approximation for the exchange-correlation functional, without any non-local, vdW correction. The choice is dictated by the fact that in  isolated, covalently bonded 2D monolayers  van der Waals interactions should be contributing only marginally to the equilibrium geometry of the structure. In Fig.~\ref{fig:DShist} we report the variation of the primitive cell area during the variable cell optimization computed as the difference between the fully optimized structure and the structure as extracted from the 3D parent. Overall, the surface variations are distributed according to a bimodal distribution, with a first peak corresponding to nearly zero variation and a second one around an expansion of 3\% while a negligible fraction of materials experienced a variation larger than 10\%. Distinguishing the data according to the functional used to optimize the 3D bulk parent (vdw DF2-c09 for all the new structures unraveled in this work, and vdw DF2-c09 or rVV10 for the ones obtained in Ref.\cite{Mounet2018}), suggests that the choice of the functional used for the bulk has a distinguishable effect on the surface area differences, with 2D structures extracted from parents optimized with vdW DF2-c09 showing on average a larger expansion during the 2D optimization. This is largely ascribed to the intrinsic differences between the PBE functional and vdW DF2-c09 with the first one giving on average an equilibrium surface area 2.5\% larger than the latter (see S.I.). In Fig.~\ref{fig:RMShist} we report instead the distribution of the normalized root-mean-squared variation of the scaled positions, to account for the internal reorganization of the atomic positions during the optimization of the monolayers. In this case the distribution is characterized by a monotonic exponential decay with the vast majority of structures showing little or no internal rearrangement. The distribution also seems unbiased by the vdW approximation used for the bulk parent. 
Both these results suggest that, as expected, the isolated optimized monolayers retain a marked structural similarity to the materials as extracted from the layered bulk parent, reiterating their aptitude to be exfoliated.


In the heatmaps of Figs.\ref{fig:DShist}-\ref{fig:RMShist} we can observe how the binding energy of a material correlates with the structural changes found during the optimization of the isolated monolayer. As one might expect, it can be observed that on average materials with lower binding energies have a smaller probability to experience large structural variations during the cell relaxation; however, this probability does not grow linearly with the binding energy but keeps a rather constant value above 50 meV/\AA$^{2}$ after an abrupt decay. This behavior might signal a more prominent presence above 50 meV/\AA$^{2}$ of structures erroneously classified as 2D (for example structures in which a 0D component has been erroneously separated including the 0D unit as a part of the 2D material instead of being properly isolated).

\subsection{Lattice matching for lateral and vertical heterostructures}

\begin{figure}[!ht]
\centering
\centering
\subfloat[][]{
\includegraphics[angle=0, width=0.5\textwidth]{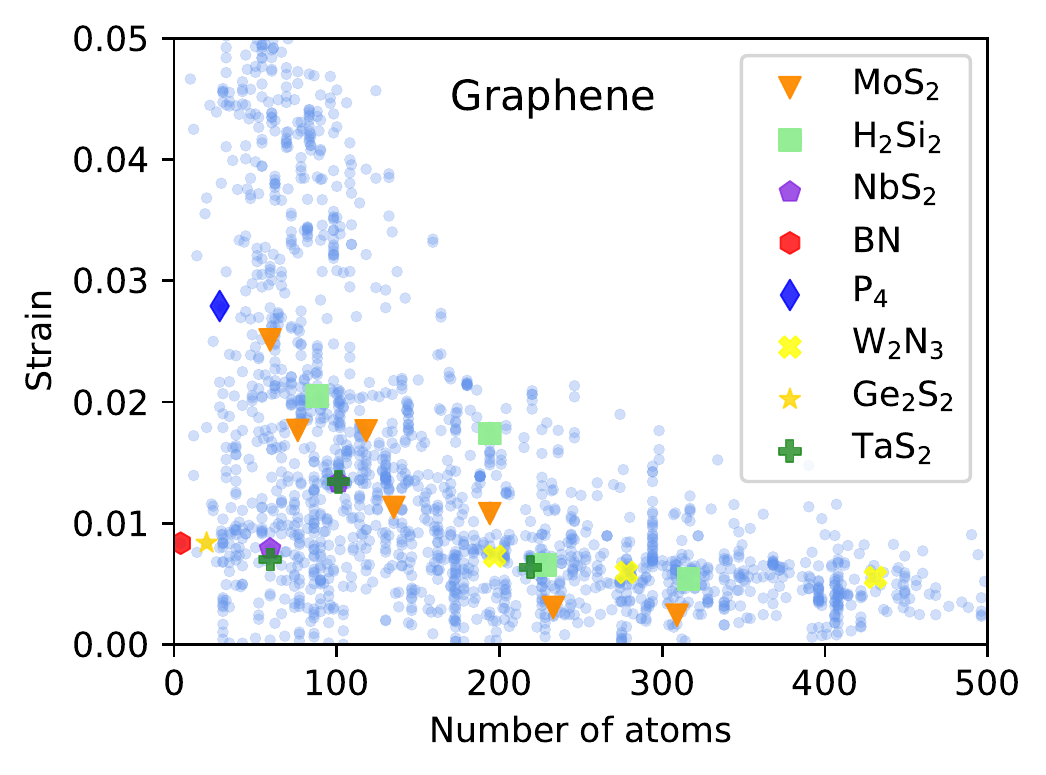}
\label{fig:Grbig}
}
\subfloat[][]{
\includegraphics[angle=0, width=0.5\textwidth]{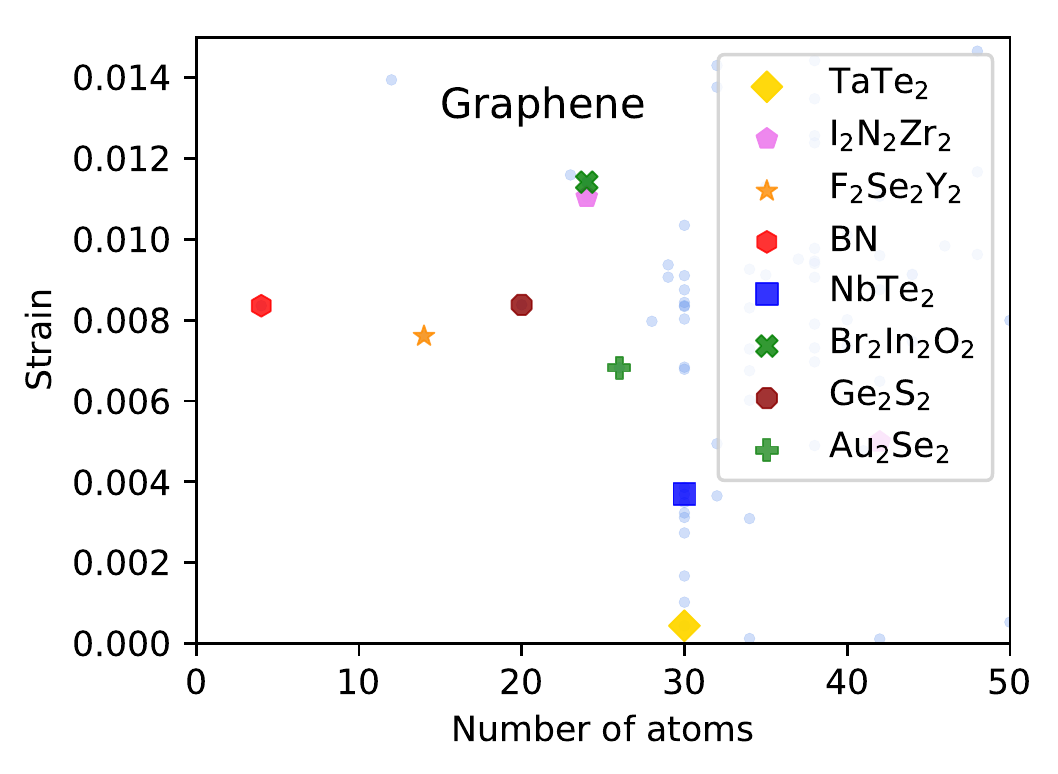}
\label{fig:Grsmall}
}\qquad
\subfloat[][]{
\includegraphics[angle=0, width=0.5\textwidth]{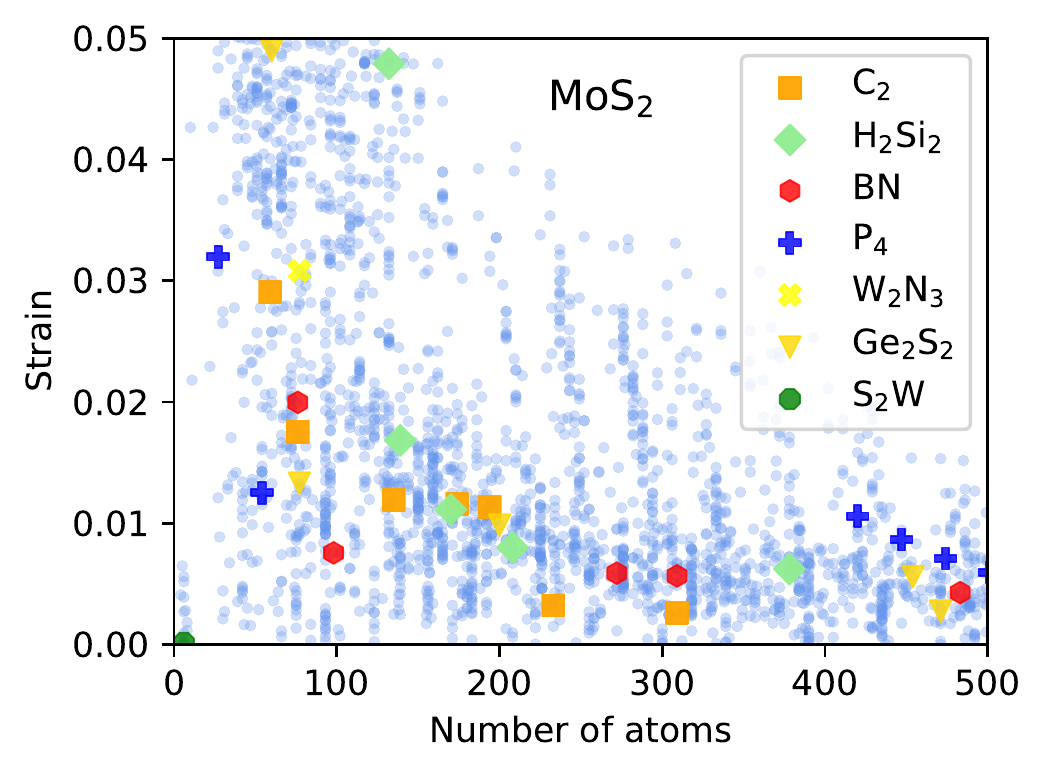}
\label{fig:mos2big}
}
\subfloat[][]{
\includegraphics[angle=0, width=0.5\textwidth]{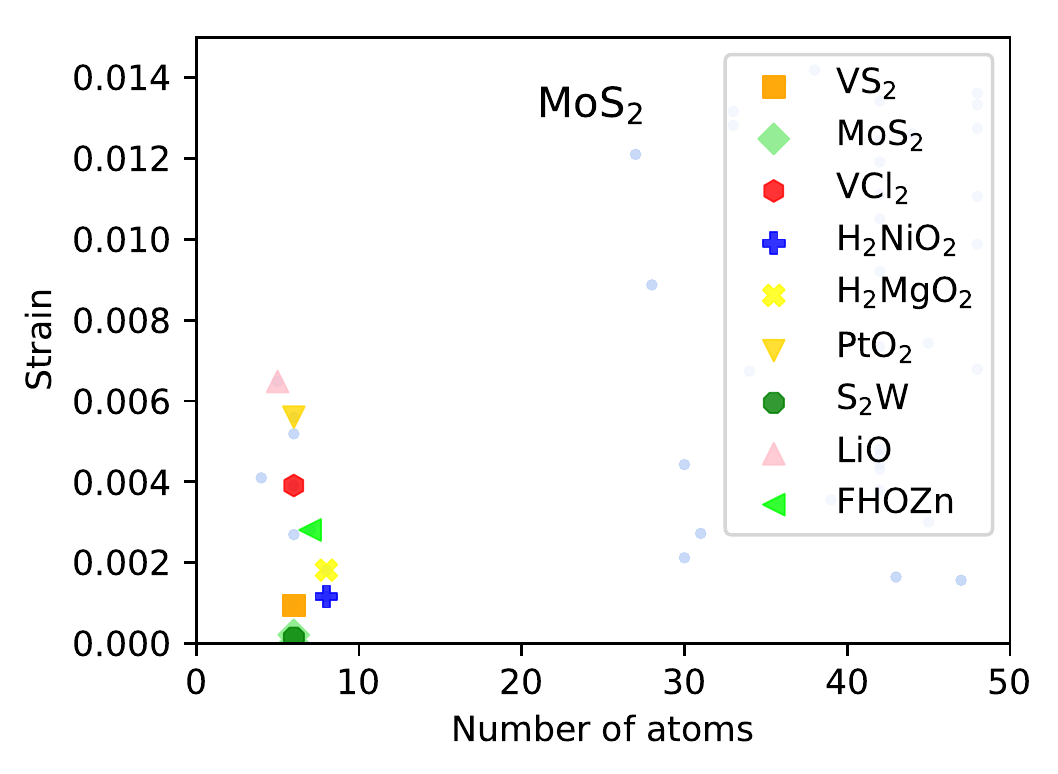}
\label{fig:mos2small}
}
\caption[]{ \subref{fig:Grbig}-\subref{fig:Grsmall} Low-strain heterostructures of graphene with materials with $\leq6$ atoms per unit cell. On the $x$-axis we report the number of atoms in the heterostructure of a certain material with graphene, and on the $y$-axis the average strain (or total deformation) applied to such a material to match the proper graphene supercell. The light blue points represent the minimum strain option for each material. Some noticeable 2D materials are highlighted as an example. \subref{fig:mos2big}-\subref{fig:mos2small} Possible low strain heterostructures of MoS$_2$ under the same conditions. 
}
\label{fig:heteros}
\end{figure}

Thanks to the progress in synthesis techniques, highly ordered lateral heterostructures have recently been realized, especially in transition-metal dicalcogenides and III-IV 2D semiconductors. These structures could enable the realization of a wide range of applications from field-effect transistors to electronic oscillators, non-volatile memory elements and plamoincs\cite{MYLi2016,CKai2018,Castellanos-Gomez2022}, but they require a nearly perfect lattice match between the two components in order to have atomically ordered defect-free interfaces. Moreover, although lattice  matching is not a strict necessity in the realization of vertically stacked heterostructures, the stacking order and possible strain effects could significantly change the electronic properties of 2D materials\cite{vanderZande2014}, especially when a 2D material is used as substrate for vdW epitaxial growth of another\cite{Shi2012}. For example, 
heterostructures formed by lattice-mismatched materials could lead to an inhomogeneous electron distribution and surface distortion that could negatively impact materials properties such as the carrier mobility\cite{Iannaccone2018}. Finally, theoretical calculations, usually performed in periodic boundary conditions, struggle with the incommensurability of target layered systems. A compromise between a favorable unstrained stacking and system size of the unit cell that can accommodate both materials is often necessary in order to lower the computational cost; pairs of materials that could form lattice matched heterostrucutres with small supercells and low strain become then especially appealing. 
For these reasons, several works already tackled the problem of searching for optimally matched heterostructures \cite{Lazic2015, Koda2016} or used such methods to study experimentally observed heterostructures. 



In this work, following the approach proposed in Ref.\cite{Lazic2015}  and on the basis of the aforementioned optimized lattice parameters, for each monolayer with up to 6 atoms per unit cell we search, within our database, for ideal materials that could give rise to lattice-matched vertical heterostructures. In Fig.~\ref{fig:heteros} we report two examples of this search for two well known materials, graphene and 2H-MoS$_2$. In the figure we report the supercell size that each material (with up to 6 atoms in the unit cell) would need to form a commensurate heterostructure with a properly sized supercell of either graphene or 2H-MoS$_2$ together with the average unit cell strain (see Ref.\cite{Lazic2015} and references therein) necessary to achieve a perfect lattice match.  Unsurprisingly, for graphene, boron nitride represents an excellent option but also Ge$_2$Se$_2$ or less known materials like F$_2$Se$_2$Y$_2$ can form lattice matched heterostructures with small unit cells and relatively small strain, while for example TaTe$_2$ can form a 30-atom heterostructure with almost zero strain.  MoS$_2$ instead offers many more options for small supercells; in  particular it is interesting to notice how, beside the well known 1T phase of MoS$_2$, and WS$_2$, materials belonging to other structural prototypes like H$_2$NiO$_2$ or H$_2$MgO$_2$ can give rise to nearly perfectly matched ($1{\times}1$) heterostructures. Tables and lists for all the 606 materials studied can be found in the S.I.; we hope this could represent a useful resource in the building of stacked heterostructures.

\subsection{Electronic properties of the 2D monolayers}
\begin{figure}[h!]
 \includegraphics[width=.5\textwidth]{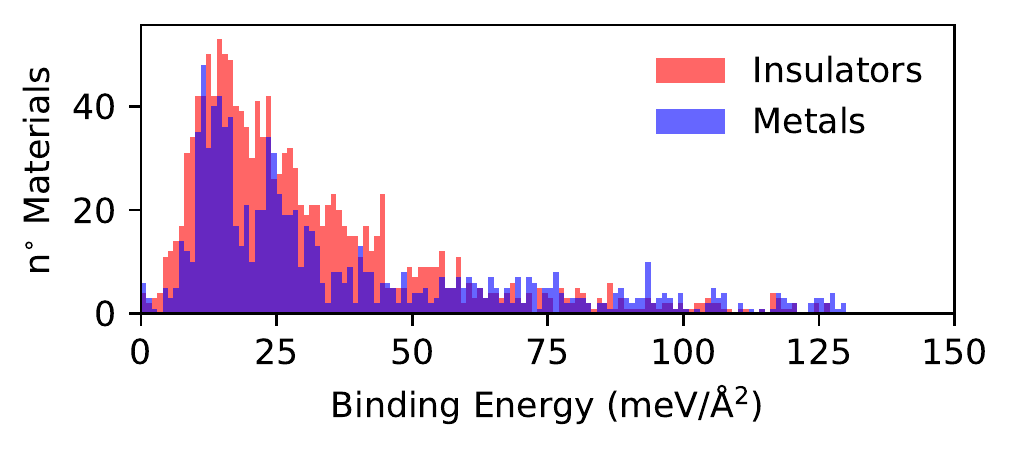}
 \caption{Distribution of the binding energies in the entire database of 3077 materials  distinguished according to the electronic properties of the monolayer. Insulators are reported in red while metals in blue.\label{fig:begap}}
\end{figure}

\begin{figure}[h!]
 \includegraphics[width=.5\textwidth]{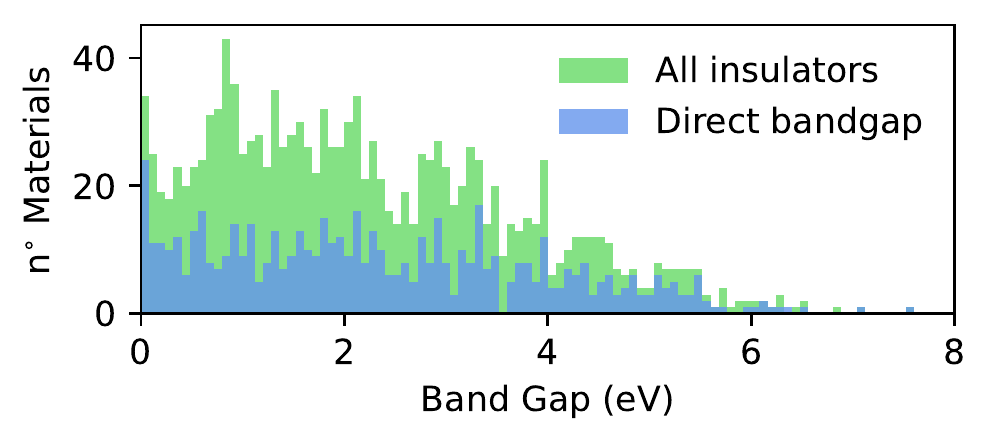}
 \caption{Distribution of the fundamental band gap at the PBE level for the semiconducting and insulating materials present in the database  obtained on the basis of the band structures computed along high-symmetry paths. In blue we highlight the contribution of  direct bandgap materials.\label{fig:gaps}}
\end{figure}

The structural optimization proved to be successful for a total of $2742$ materials while $335$ failed to converge. For these $2742$ materials we compute band structures at the PBE level along high-symmetry paths. Magnetism is not considered and all the structures have been treated as non magnetic. We find that 65\% of the materials are metallic while the rest is insulating or semiconducting. In Fig.~\ref{fig:begap} we report the distribution of the binding energies for the entire database, classified according to the electronic properties of the monolayer. We can note that at very low binding energies both metallic and semiconducting/insulating materials are distributed in a relatively similar way, but at binding energies between 25 and 50 meV/\AA$^{2}$ the presence of insulating compounds is marginally higher. On the contrary for higher binding energies (above 50 meV) the presence of metallic compounds becomes more dominant, possibly signaling again a higher presence of materials with unsaturated bonds. In Fig. \ref{fig:gaps} we report instead the distribution of the fundamental band gaps at the PBE level for the insulating and semiconducting monolayers. The band gaps follow a quite broad distribution with a peak around 1 eV, followed by a wide plateau in the 1 to 2.5 eV energy range, and a subsequent slow decay. In the picture we also highlight the fraction of direct bandgap materials, particularly relevant for optical and optoelectronic applications. These materials are distributed fairly uniformly throughout the energy range accounting for roughly one third of the materials for moderate bandgaps but with ratios increasing up to one half for extremely small or very large gaps. Interestingly, if one accounts for the well known underestimation of the gap with DFT, materials with bandgaps around 1 eV would tend to be shifted towards higher values offering hundreds of candidates in the optimal range for solar light harvesting.


\begin{figure}[h!]
\includegraphics[width=0.6\textwidth]{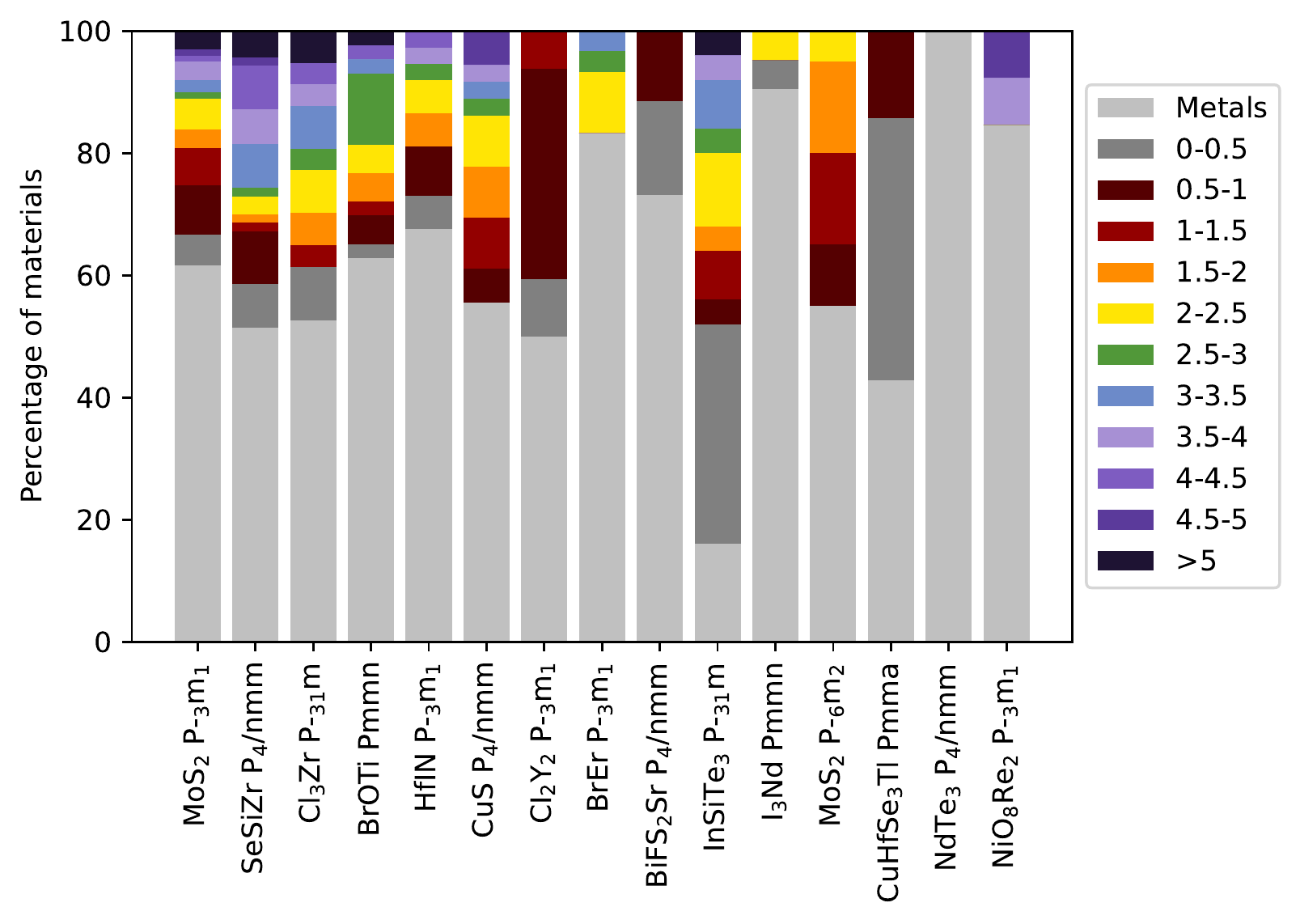}
\caption{Distribution of the PBE band gaps for the materials in the fifteen most common prototype classes. The $0{-}5$ eV range has been equally divided in ten parts each represented by different colors.}
\label{fig:protoGAP}
\end{figure}

Finally, in Fig.~\ref{fig:protoGAP} we show the distribution of the PBE fundamental bandgaps within each of the prototypical families identified in Fig.~\ref{fig:2Dprototypes}. It can be noted that while certain classes like the tri-chalcogenides and tri-halides represented respectively by NdTe$_3$ and NdI$_3$ are almost entirely dominated by metallic compounds or at most small bandgap materials,  (e.g. Cl$_2$Y$_2$, BiFS$_2$Sr and CuHfSe$_3$Tl),  others offer a broad range of properties ranging from metallic to wide bandgaps.  This classification could be used as a guide in a combinatorial search through atomic substitutions for specific applications, or to highlight the potential of bandgap tuning through atomic alloying within each class. 


\FloatBarrier
\subsection{Large bandgap 2D insulators}

At the extreme edge of the distribution in Fig.~\ref{fig:gaps} we can observe few materials with large bandgaps that could be interesting as insulating layers in nanoscale electronic devices.  In a recent work\cite{Knobloch2021} it has been systematically shown how properties such as the bandgap, the dielectric constant and the effective masses of the insulating capping layer could influence the performance of ultrascaled tunneling transistors realized with 2D channels, especially  through their influence on the leakage current and carrier mobility. In particular, materials able to form a clean van der Waals interface with a large bandgap (approximately around 12 eV), with large effective masses and with a static dielectric constant of 10 or above would be the ideal candidates to realize this insulating layer. Unfortunately materials with such characteristics are extremely rare. With this application in mind, here we study in further detail the materials that present the largest bandgaps at the DFT level in our screening, namely B$_2$F$_8$Na (7.5 eV), F$_2$H$_6$O$_2$ (7.1 eV) and B$_2$O$_9$S$_2$ (6.9 eV). All these materials are dynamically stable as proven by their phonon dispersions (see S.I.), and have dielectric constants of 4.7, 5.8, 5.15 respectively (comparable or slightly higher than h-BN, 4.7) and significantly larger bandgaps. Moreover, they have only slightly larger effective masses ($m^{*}_{e1}{=}0.65$,  $m^{*}_{e2}{=}0.64$), ($m^{*}_{e1}{=}0.57$,  $m^{*}_{e2}{=}0.55$), $m^{*}_{e1}{=}0.64$,  $m^{*}_{e2}{=}0.52$) respectively for the electrons, but extremely large and anisotropic effective masses for the holes ($m^{*}_{h1}{=}17.7$,  $m^{*}_{h2}{=}34.1$), ($m^{*}_{h1}{=}5.38$,  $m^{*}_{h2}{=}47.9$), ($m^{*}_{h1}{=}2.86$,  $m^{*}_{h2}{=}7.69$) compared to 2D h-BN ($m^{*}{=}0.54$ for both holes and electrons in all directions). These figures indicate that all the proposed materials should outperform h-BN as insulating layers while still maintaining the possibility of forming ideal vdW interfaces, and they thus deserve more in-depth experimental studies to assess their performance in operational conditions. 


\section{Conclusions}
In conclusion, in this work we screened for novel two dimensional monolayers a database of experimental structures (MPDS) as well as the most up-to-date versions of the two experimental databases already included in our previous study. This work has led to the discovery of 1252 new unique monolayers, bringing the total to 3077 compounds and, notably, almost doubling the number of easily exfoliable materials (2004). Moreover, we optimized the structural properties of each material treated as an isolated monolayer and studied their electronic properties. On the basis of their optimized geometries, for the subset materials with up to 6 atoms in the unit cell, we suggest all possible combinations of materials that can give rise to lattice-matched lateral or vertical heterostructures.
We computed and analyzed the electronic band structure at the PBE level for each compound and studied the distribution of the fundamental electronic properties throughout the database and within the most common prototype classes. Finally, we highlighted a handful of large-gap insulating materials that could possibly outperform boron nitride as insulating layers for ultrascaled transistors. All the information about the materials and properties of this study are presented in the Supplementary Material and are available on the Materials Cloud in the Discover, Explore and Archive sections for the benefit of the experimental and theoretical community.

\begin{acknowledgement}
The authors acknowledge the help of Dr.~Sebaastian P. Huber in the process of extracting the data form online database and import them into AiiDA, Dr.~Davide Grassano and Dr.~Elsa Passaro for the preparation of the Materials Cloud explore section.
This work was supported by the Center for Computational Design and Discovery on Novel Materials NCCR MARVEL of the Swiss National Science Foundation.  D.C.\ also acknowledges the support from the EPFL Fellows fellowship programme co-funded by Marie Sklodowska-Curie, Horizon 2020 grant agreement no. 665667. M.G.\ acknowledge the support of the Italian Ministry for University and Research through the Levi-Montalcini program.
Simulation time was awarded by CSCS on Piz Daint (production project s825) and by PRACE on  Marconi at Cineca, Italy (project id.\ 2016163963).
\end{acknowledgement}

\clearpage
\bibliography{biblio}

\providecommand{\latin}[1]{#1}
\makeatletter
\providecommand{\doi}
  {\begingroup\let\do\@makeother\dospecials
  \catcode`\{=1 \catcode`\}=2 \doi@aux}
\providecommand{\doi@aux}[1]{\endgroup\texttt{#1}}
\makeatother
\providecommand*\mcitethebibliography{\thebibliography}
\csname @ifundefined\endcsname{endmcitethebibliography}
  {\let\endmcitethebibliography\endthebibliography}{}
\begin{mcitethebibliography}{57}
\providecommand*\natexlab[1]{#1}
\providecommand*\mciteSetBstSublistMode[1]{}
\providecommand*\mciteSetBstMaxWidthForm[2]{}
\providecommand*\mciteBstWouldAddEndPuncttrue
  {\def\EndOfBibitem{\unskip.}}
\providecommand*\mciteBstWouldAddEndPunctfalse
  {\let\EndOfBibitem\relax}
\providecommand*\mciteSetBstMidEndSepPunct[3]{}
\providecommand*\mciteSetBstSublistLabelBeginEnd[3]{}
\providecommand*\EndOfBibitem{}
\mciteSetBstSublistMode{f}
\mciteSetBstMaxWidthForm{subitem}{(\alph{mcitesubitemcount})}
\mciteSetBstSublistLabelBeginEnd
  {\mcitemaxwidthsubitemform\space}
  {\relax}
  {\relax}

\bibitem[Radisavljevic \latin{et~al.}(2011)Radisavljevic, Radenovic, Brivio,
  Giacometti, and Kis]{Radisavljevic:2011}
Radisavljevic,~B.; Radenovic,~A.; Brivio,~J.; Giacometti,~V.; Kis,~A.
  {Single-layer MoS$_2$ transistors}. \emph{Nature Nanotechnology}
  \textbf{2011}, \emph{6}, 147--150\relax
\mciteBstWouldAddEndPuncttrue
\mciteSetBstMidEndSepPunct{\mcitedefaultmidpunct}
{\mcitedefaultendpunct}{\mcitedefaultseppunct}\relax
\EndOfBibitem
\bibitem[Wang \latin{et~al.}(2012)Wang, Kalantar-Zadeh, Kis, Coleman, and
  Strano]{Wang2012}
Wang,~Q.~H.; Kalantar-Zadeh,~K.; Kis,~A.; Coleman,~J.~N.; Strano,~M.~S.
  Electronics and optoelectronics of two-dimensional transition metal
  dichalcogenides. \emph{Nat. Nanotech.} \textbf{2012}, \emph{7},
  699--712\relax
\mciteBstWouldAddEndPuncttrue
\mciteSetBstMidEndSepPunct{\mcitedefaultmidpunct}
{\mcitedefaultendpunct}{\mcitedefaultseppunct}\relax
\EndOfBibitem
\bibitem[Liu \latin{et~al.}(2014)Liu, Neal, Zhu, Luo, Xu, Tom{\~A}{!'}nek, and
  Ye]{Liu:2014}
Liu,~H.; Neal,~A.~T.; Zhu,~Z.; Luo,~Z.; Xu,~X.; Tom{\~A}{!'}nek,~D.; Ye,~P.~D.
  Phosphorene: An Unexplored {2D} Semiconductor with a High Hole Mobility.
  \emph{ACS Nano} \textbf{2014}, \emph{8}, 4033--4041\relax
\mciteBstWouldAddEndPuncttrue
\mciteSetBstMidEndSepPunct{\mcitedefaultmidpunct}
{\mcitedefaultendpunct}{\mcitedefaultseppunct}\relax
\EndOfBibitem
\bibitem[Roy \latin{et~al.}(2014)Roy, Tosun, Kang, Sachid, Desai, Hettick, Hu,
  and Javey]{Roy2014}
Roy,~T.; Tosun,~M.; Kang,~J.~S.; Sachid,~A.~B.; Desai,~S.~B.; Hettick,~M.;
  Hu,~C.~C.; Javey,~A. Field-Effect Transistors Built from All Two-Dimensional
  Material Components. \emph{ACS Nano} \textbf{2014}, \emph{8},
  6259--6264\relax
\mciteBstWouldAddEndPuncttrue
\mciteSetBstMidEndSepPunct{\mcitedefaultmidpunct}
{\mcitedefaultendpunct}{\mcitedefaultseppunct}\relax
\EndOfBibitem
\bibitem[Chhowalla \latin{et~al.}(2016)Chhowalla, Jena, and
  Zhang]{Chhowalla2016}
Chhowalla,~M.; Jena,~D.; Zhang,~H. Two-dimensional semiconductors for
  transistors. \emph{Nature Reviews Materials} \textbf{2016}, \emph{1}, 16052,
  Review Article\relax
\mciteBstWouldAddEndPuncttrue
\mciteSetBstMidEndSepPunct{\mcitedefaultmidpunct}
{\mcitedefaultendpunct}{\mcitedefaultseppunct}\relax
\EndOfBibitem
\bibitem[Klinkert \latin{et~al.}(2020)Klinkert, Szab{\'o}, Stieger, Campi,
  Marzari, and Luisier]{Klinkert2020}
Klinkert,~C.; Szab{\'o},~{\'A}.; Stieger,~C.; Campi,~D.; Marzari,~N.;
  Luisier,~M. 2-D Materials for Ultrascaled Field-Effect Transistors: One
  Hundred Candidates under the Ab Initio Microscope. \emph{ACS Nano}
  \textbf{2020}, \emph{14}, 8605--8615\relax
\mciteBstWouldAddEndPuncttrue
\mciteSetBstMidEndSepPunct{\mcitedefaultmidpunct}
{\mcitedefaultendpunct}{\mcitedefaultseppunct}\relax
\EndOfBibitem
\bibitem[Sun \latin{et~al.}(2012)Sun, Cheng, Gao, Sun, Liu, Liu, Lei, Yao, and
  He]{He2012}
Sun,~Y.; Cheng,~H.; Gao,~S.; Sun,~Z.; Liu,~Q.; Liu,~Q.; Lei,~F.; Yao,~T.;
  He,~J. S.~W. {Freestanding Tin Disulfide Single-Layers Realizing Efficient
  Visible-Light Water Splitting.} \emph{Angew. Chem., Int. Ed.} \textbf{2012},
  \emph{51}, 8727--8731\relax
\mciteBstWouldAddEndPuncttrue
\mciteSetBstMidEndSepPunct{\mcitedefaultmidpunct}
{\mcitedefaultendpunct}{\mcitedefaultseppunct}\relax
\EndOfBibitem
\bibitem[Sun \latin{et~al.}(2012)Sun, Sun, Gao, Cheng, Liu, Piao, Yao, Wu, and
  Hu]{Sun2012}
Sun,~Y.; Sun,~Z.; Gao,~S.; Cheng,~H.; Liu,~Q.; Piao,~J.; Yao,~T.; Wu,~C.;
  Hu,~S. S.~W. {Fabrication of Flexible and Freestanding Zinc Chalcogenide
  Single Layers}. \emph{Nat. Commun.} \textbf{2012}, \emph{3}, 1057\relax
\mciteBstWouldAddEndPuncttrue
\mciteSetBstMidEndSepPunct{\mcitedefaultmidpunct}
{\mcitedefaultendpunct}{\mcitedefaultseppunct}\relax
\EndOfBibitem
\bibitem[Qian \latin{et~al.}(2014)Qian, Liu, Fu, and Li]{Qian2016}
Qian,~X.; Liu,~J.; Fu,~L.; Li,~J. Quantum spin Hall effect in two-dimensional
  transition metal dichalcogenides. \emph{Science} \textbf{2014}, \emph{346},
  1344--1347\relax
\mciteBstWouldAddEndPuncttrue
\mciteSetBstMidEndSepPunct{\mcitedefaultmidpunct}
{\mcitedefaultendpunct}{\mcitedefaultseppunct}\relax
\EndOfBibitem
\bibitem[Marrazzo \latin{et~al.}(2018)Marrazzo, Gibertini, Campi, Mounet, and
  Marzari]{Marrazzo2018}
Marrazzo,~A.; Gibertini,~M.; Campi,~D.; Mounet,~N.; Marzari,~N. Prediction of a
  Large-Gap and Switchable Kane-Mele Quantum Spin Hall Insulator. \emph{Phys.
  Rev. Lett.} \textbf{2018}, \emph{120}, 117701\relax
\mciteBstWouldAddEndPuncttrue
\mciteSetBstMidEndSepPunct{\mcitedefaultmidpunct}
{\mcitedefaultendpunct}{\mcitedefaultseppunct}\relax
\EndOfBibitem
\bibitem[Campi \latin{et~al.}(2021)Campi, Kumari, and Marzari]{Campi2021}
Campi,~D.; Kumari,~S.; Marzari,~N. Prediction of Phonon-Mediated
  Superconductivity with High Critical Temperature in the Two-Dimensional
  Topological Semimetal W2N3. \emph{Nano Letters} \textbf{2021}, \emph{21},
  3435--3442\relax
\mciteBstWouldAddEndPuncttrue
\mciteSetBstMidEndSepPunct{\mcitedefaultmidpunct}
{\mcitedefaultendpunct}{\mcitedefaultseppunct}\relax
\EndOfBibitem
\bibitem[Geim and Grigorieva(2013)Geim, and Grigorieva]{geim2013van}
Geim,~A.; Grigorieva,~I. Van der {Waals} heterostructures. \emph{Nature}
  \textbf{2013}, \emph{499}, 419--425\relax
\mciteBstWouldAddEndPuncttrue
\mciteSetBstMidEndSepPunct{\mcitedefaultmidpunct}
{\mcitedefaultendpunct}{\mcitedefaultseppunct}\relax
\EndOfBibitem
\bibitem[Backes \latin{et~al.}(2019)Backes, Campi, Szydlowska, Synnatschke,
  Ojala, Rashvand, Harvey, Griffin, Sofer, Marzari, Coleman, and
  O'Regan]{Backes2019}
Backes,~C.; Campi,~D.; Szydlowska,~B.~M.; Synnatschke,~K.; Ojala,~E.;
  Rashvand,~F.; Harvey,~A.; Griffin,~A.; Sofer,~Z.; Marzari,~N.;
  Coleman,~J.~N.; O'Regan,~D.~D. Equipartition of Energy Defines the
  Size--Thickness Relationship in Liquid-Exfoliated Nanosheets. \emph{ACS Nano}
  \textbf{2019}, \emph{13}, 7050--7061\relax
\mciteBstWouldAddEndPuncttrue
\mciteSetBstMidEndSepPunct{\mcitedefaultmidpunct}
{\mcitedefaultendpunct}{\mcitedefaultseppunct}\relax
\EndOfBibitem
\bibitem[Huang \latin{et~al.}(2020)Huang, Pan, Yang, Bao, Meng, Luo, Cai, Liu,
  Zhao, Zhou, Wu, Zhu, Huang, Liu, Liu, Cheng, Wu, Tian, Gu, Shi, Guo, Cheng,
  Hu, Zhao, Yang, Sutter, Sutter, Wang, Ji, Zhou, and Gao]{Huang2020}
Huang,~Y. \latin{et~al.}  Author Correction: Universal mechanical exfoliation
  of large-area 2D crystals. \emph{Nature Communications} \textbf{2020},
  \emph{11}, 2938\relax
\mciteBstWouldAddEndPuncttrue
\mciteSetBstMidEndSepPunct{\mcitedefaultmidpunct}
{\mcitedefaultendpunct}{\mcitedefaultseppunct}\relax
\EndOfBibitem
\bibitem[Gould \latin{et~al.}(2016)Gould, Leb{\`e}gue, Bj{\"o}rkman, and
  Dobson]{Gould2016}
Gould,~T.; Leb{\`e}gue,~S.; Bj{\"o}rkman,~T.; Dobson,~J. In \emph{2D
  Materials}; Iacopi,~F., Boeckl,~J.~J., Jagadish,~C., Eds.; Semiconductors and
  Semimetals; Elsevier, 2016; Vol.~95; Chapter 1, pp 1 -- 33\relax
\mciteBstWouldAddEndPuncttrue
\mciteSetBstMidEndSepPunct{\mcitedefaultmidpunct}
{\mcitedefaultendpunct}{\mcitedefaultseppunct}\relax
\EndOfBibitem
\bibitem[Leb{\`e}gue \latin{et~al.}(2013)Leb{\`e}gue, Bj{\"o}rkman,
  Klintenberg, Nieminen, and Eriksson]{Nieminen}
Leb{\`e}gue,~S.; Bj{\"o}rkman,~T.; Klintenberg,~M.; Nieminen,~R.~M.;
  Eriksson,~O. Two-Dimensional Materials from Data Filtering and Ab Initio
  Calculations. \emph{Physical Review X} \textbf{2013}, \emph{3}, 031002\relax
\mciteBstWouldAddEndPuncttrue
\mciteSetBstMidEndSepPunct{\mcitedefaultmidpunct}
{\mcitedefaultendpunct}{\mcitedefaultseppunct}\relax
\EndOfBibitem
\bibitem[Rasmussen and Thygesen(2015)Rasmussen, and
  Thygesen]{rasmussen2015computational}
Rasmussen,~F.~A.; Thygesen,~K.~S. Computational {2D Materials Database}:
  {Electronic} Structure of Transition-Metal Dichalcogenides and Oxides.
  \emph{The Journal of Physical Chemistry C} \textbf{2015}, \emph{119},
  13169--13183\relax
\mciteBstWouldAddEndPuncttrue
\mciteSetBstMidEndSepPunct{\mcitedefaultmidpunct}
{\mcitedefaultendpunct}{\mcitedefaultseppunct}\relax
\EndOfBibitem
\bibitem[Choudhary \latin{et~al.}(2017)Choudhary, Kalish, Beams, and
  Tavazza]{Choudhary2017}
Choudhary,~K.; Kalish,~I.; Beams,~R.; Tavazza,~F. High-throughput
  Identification and Characterization of Two-dimensional Materials using
  Density functional theory. \emph{Scientific Reports} \textbf{2017}, \emph{7},
  5179\relax
\mciteBstWouldAddEndPuncttrue
\mciteSetBstMidEndSepPunct{\mcitedefaultmidpunct}
{\mcitedefaultendpunct}{\mcitedefaultseppunct}\relax
\EndOfBibitem
\bibitem[Ashton \latin{et~al.}(2017)Ashton, Paul, Sinnott, and
  Hennig]{Ashton2017}
Ashton,~M.; Paul,~J.; Sinnott,~S.~B.; Hennig,~R.~G. Topology-Scaling
  Identification of Layered Solids and Stable Exfoliated 2D Materials.
  \emph{Phys. Rev. Lett.} \textbf{2017}, \emph{118}, 106101\relax
\mciteBstWouldAddEndPuncttrue
\mciteSetBstMidEndSepPunct{\mcitedefaultmidpunct}
{\mcitedefaultendpunct}{\mcitedefaultseppunct}\relax
\EndOfBibitem
\bibitem[Cheon \latin{et~al.}(2017)Cheon, Duerloo, Sendek, Porter, Chen, and
  Reed]{Cheon2017}
Cheon,~G.; Duerloo,~K.-A.~N.; Sendek,~A.~D.; Porter,~C.; Chen,~Y.; Reed,~E.~J.
  Data Mining for New Two- and One-Dimensional Weakly Bonded Solids and
  Lattice-Commensurate Heterostructures. \emph{Nano Letters} \textbf{2017},
  \emph{17}, 1915--1923\relax
\mciteBstWouldAddEndPuncttrue
\mciteSetBstMidEndSepPunct{\mcitedefaultmidpunct}
{\mcitedefaultendpunct}{\mcitedefaultseppunct}\relax
\EndOfBibitem
\bibitem[Mounet \latin{et~al.}(2018)Mounet, Gibertini, Schwaller, Campi,
  Merkys, Marrazzo, Thib, Castelli, Cepellotti, Pizzi, and Marzari]{Mounet2018}
Mounet,~N.; Gibertini,~M.; Schwaller,~P.; Campi,~D.; Merkys,~A.; Marrazzo,~A.;
  Thib,; Castelli,~I.~E.; Cepellotti,~A.; Pizzi,~G.; Marzari,~N.
  Two-dimensional materials from high-throughput computational exfoliation of
  experimentally known compounds. \emph{Nature Nanotechnology} \textbf{2018},
  \emph{13}, 246\relax
\mciteBstWouldAddEndPuncttrue
\mciteSetBstMidEndSepPunct{\mcitedefaultmidpunct}
{\mcitedefaultendpunct}{\mcitedefaultseppunct}\relax
\EndOfBibitem
\bibitem[Haastrup \latin{et~al.}(2018)Haastrup, Strange, Pandey, Deilmann,
  Schmidt, Hinsche, Gjerding, Torelli, Larsen, Riis-Jensen, Gath, Jacobsen,
  Mortensen, Olsen, and Thygesen]{Haastrup2018}
Haastrup,~S.; Strange,~M.; Pandey,~M.; Deilmann,~T.; Schmidt,~P.~S.;
  Hinsche,~N.~F.; Gjerding,~M.~N.; Torelli,~D.; Larsen,~P.~M.;
  Riis-Jensen,~A.~C.; Gath,~J.; Jacobsen,~K.~W.; Mortensen,~J.~J.; Olsen,~T.;
  Thygesen,~K.~S. The Computational 2D Materials Database: high-throughput
  modeling and discovery of atomically thin crystals. \emph{2D Materials}
  \textbf{2018}, \emph{5}, 042002\relax
\mciteBstWouldAddEndPuncttrue
\mciteSetBstMidEndSepPunct{\mcitedefaultmidpunct}
{\mcitedefaultendpunct}{\mcitedefaultseppunct}\relax
\EndOfBibitem
\bibitem[Zhou \latin{et~al.}(2019)Zhou, Shen, Costa, Persson, Ong, Huck, Lu,
  Ma, Chen, Tang, and Feng]{Zhou2019}
Zhou,~J.; Shen,~L.; Costa,~M.~D.; Persson,~K.~A.; Ong,~S.~P.; Huck,~P.; Lu,~Y.;
  Ma,~X.; Chen,~Y.; Tang,~H.; Feng,~Y.~P. 2DMatPedia, an open computational
  database of two-dimensional materials from top-down and bottom-up approaches.
  \emph{Scientific Data} \textbf{2019}, \emph{6}, 86\relax
\mciteBstWouldAddEndPuncttrue
\mciteSetBstMidEndSepPunct{\mcitedefaultmidpunct}
{\mcitedefaultendpunct}{\mcitedefaultseppunct}\relax
\EndOfBibitem
\bibitem[ICS()]{ICSD}
{Inorganic Crystal Structure Database (ICSD)}.
  http://www.fiz-karlsruhe.com/icsd.html\relax
\mciteBstWouldAddEndPuncttrue
\mciteSetBstMidEndSepPunct{\mcitedefaultmidpunct}
{\mcitedefaultendpunct}{\mcitedefaultseppunct}\relax
\EndOfBibitem
\bibitem[Gra\v{z}ulis \latin{et~al.}(2012)Gra\v{z}ulis, Da\v{s}kevi\v{c},
  Merkys, Chateigner, Lutterotti, Quir\'os, Serebryanaya, Moeck, Downs, and
  Le~Bail]{COD}
Gra\v{z}ulis,~S.; Da\v{s}kevi\v{c},~A.; Merkys,~A.; Chateigner,~D.;
  Lutterotti,~L.; Quir\'os,~M.; Serebryanaya,~N.~R.; Moeck,~P.; Downs,~R.~T.;
  Le~Bail,~A. Crystallography Open Database ({COD}): an open-access collection
  of crystal structures and platform for world-wide collaboration.
  \emph{Nucleic Acids Research} \textbf{2012}, \emph{40}, D420--D427\relax
\mciteBstWouldAddEndPuncttrue
\mciteSetBstMidEndSepPunct{\mcitedefaultmidpunct}
{\mcitedefaultendpunct}{\mcitedefaultseppunct}\relax
\EndOfBibitem
\bibitem[Novoselov \latin{et~al.}(2004)Novoselov, Geim, Morozov, Jiang, Zhang,
  Dubonos, Grigorieva, and Firsov]{Novoselov2004}
Novoselov,~K.~S.; Geim,~A.~K.; Morozov,~S.~V.; Jiang,~D.; Zhang,~Y.;
  Dubonos,~S.~V.; Grigorieva,~I.~V.; Firsov,~A.~A. Electric Field Effect in
  Atomically Thin Carbon Films. \emph{Science} \textbf{2004}, \emph{306},
  666--669\relax
\mciteBstWouldAddEndPuncttrue
\mciteSetBstMidEndSepPunct{\mcitedefaultmidpunct}
{\mcitedefaultendpunct}{\mcitedefaultseppunct}\relax
\EndOfBibitem
\bibitem[Coleman \latin{et~al.}(2011)Coleman, Lotya, O{\textquoteright}Neill,
  Bergin, King, Khan, Young, Gaucher, De, Smith, Shvets, Arora, Stanton, Kim,
  Lee, Kim, Duesberg, Hallam, Boland, Wang, Donegan, Grunlan, Moriarty,
  Shmeliov, Nicholls, Perkins, Grieveson, Theuwissen, McComb, Nellist, and
  Nicolosi]{Coleman2011}
Coleman,~J.~N. \latin{et~al.}  Two-Dimensional Nanosheets Produced by Liquid
  Exfoliation of Layered Materials. \emph{Science} \textbf{2011}, \emph{331},
  568--571\relax
\mciteBstWouldAddEndPuncttrue
\mciteSetBstMidEndSepPunct{\mcitedefaultmidpunct}
{\mcitedefaultendpunct}{\mcitedefaultseppunct}\relax
\EndOfBibitem
\bibitem[Villars \latin{et~al.}(1998)Villars, Onodera, and Iwata]{Villars:1998}
Villars,~P.; Onodera,~N.; Iwata,~S. {The Linus Pauling file (LPF) and its
  application to materials design}. \emph{Journal of Alloys and Compounds}
  \textbf{1998}, \emph{279}, 1--7\relax
\mciteBstWouldAddEndPuncttrue
\mciteSetBstMidEndSepPunct{\mcitedefaultmidpunct}
{\mcitedefaultendpunct}{\mcitedefaultseppunct}\relax
\EndOfBibitem
\bibitem[Knobloch \latin{et~al.}(2021)Knobloch, Illarionov, Ducry, Schleich,
  Wachter, Watanabe, Taniguchi, Mueller, Waltl, Lanza, Vexler, Luisier, and
  Grasser]{Knobloch2021}
Knobloch,~T.; Illarionov,~Y.~Y.; Ducry,~F.; Schleich,~C.; Wachter,~S.;
  Watanabe,~K.; Taniguchi,~T.; Mueller,~T.; Waltl,~M.; Lanza,~M.;
  Vexler,~M.~I.; Luisier,~M.; Grasser,~T. The performance limits of hexagonal
  boron nitride as an insulator for scaled CMOS devices based on
  two-dimensional materials. \emph{Nature Electronics} \textbf{2021}, \emph{4},
  98--108\relax
\mciteBstWouldAddEndPuncttrue
\mciteSetBstMidEndSepPunct{\mcitedefaultmidpunct}
{\mcitedefaultendpunct}{\mcitedefaultseppunct}\relax
\EndOfBibitem
\bibitem[Pizzi \latin{et~al.}(2016)Pizzi, Cepellotti, Sabatini, Marzari, and
  Kozinsky]{AiiDA}
Pizzi,~G.; Cepellotti,~A.; Sabatini,~R.; Marzari,~N.; Kozinsky,~B. {AiiDA}:
  automated interactive infrastructure and database for computational science.
  \emph{Computational Materials Science} \textbf{2016}, \emph{111}, 218 --
  230\relax
\mciteBstWouldAddEndPuncttrue
\mciteSetBstMidEndSepPunct{\mcitedefaultmidpunct}
{\mcitedefaultendpunct}{\mcitedefaultseppunct}\relax
\EndOfBibitem
\bibitem[Huber \latin{et~al.}(2020)Huber, Zoupanos, Uhrin, Talirz, Kahle,
  H{\"a}uselmann, Gresch, M{\"u}ller, Yakutovich, Andersen, Ramirez, Adorf,
  Gargiulo, Kumbhar, Passaro, Johnston, Merkys, Cepellotti, Mounet, Marzari,
  Kozinsky, and Pizzi]{Huber2020}
Huber,~S.~P. \latin{et~al.}  AiiDA 1.0, a scalable computational infrastructure
  for automated reproducible workflows and data provenance. \emph{Scientific
  Data} \textbf{2020}, \emph{7}, 300\relax
\mciteBstWouldAddEndPuncttrue
\mciteSetBstMidEndSepPunct{\mcitedefaultmidpunct}
{\mcitedefaultendpunct}{\mcitedefaultseppunct}\relax
\EndOfBibitem
\bibitem[Talirz \latin{et~al.}(2020)Talirz, Kumbhar, Passaro, Yakutovich,
  Granata, Gargiulo, Borelli, Uhrin, Huber, Zoupanos, Adorf, Andersen,
  Sch{\"u}tt, Pignedoli, Passerone, VandeVondele, Schulthess, Smit, Pizzi, and
  Marzari]{Talirz2020}
Talirz,~L. \latin{et~al.}  Materials Cloud, a platform for open computational
  science. \emph{Scientific Data} \textbf{2020}, \emph{7}, 299\relax
\mciteBstWouldAddEndPuncttrue
\mciteSetBstMidEndSepPunct{\mcitedefaultmidpunct}
{\mcitedefaultendpunct}{\mcitedefaultseppunct}\relax
\EndOfBibitem
\bibitem[Campi \latin{et~al.}(2022)Campi, Mounet, Gibertini, Pizzi, and
  Marzari]{Campi2022MC}
Campi,~D.; Mounet,~N.; Gibertini,~M.; Pizzi,~G.; Marzari,~N. The Materials
  Cloud 2D database (MC2D). \emph{Materials Cloud Archive} \textbf{2022},
  \relax
\mciteBstWouldAddEndPunctfalse
\mciteSetBstMidEndSepPunct{\mcitedefaultmidpunct}
{}{\mcitedefaultseppunct}\relax
\EndOfBibitem
\bibitem[Hall and McMahon(2005)Hall, and McMahon]{CIF}
Hall,~S., McMahon,~B., Eds. \emph{{International Tables for Crystallography}};
  Springer, 2005; Vol.~G\relax
\mciteBstWouldAddEndPuncttrue
\mciteSetBstMidEndSepPunct{\mcitedefaultmidpunct}
{\mcitedefaultendpunct}{\mcitedefaultseppunct}\relax
\EndOfBibitem
\bibitem[Ong \latin{et~al.}(2013)Ong, Richards, Jain, Hautier, Kocher, Cholia,
  Gunter, Chevrier, Persson, and Ceder]{pymatgen}
Ong,~S.~P.; Richards,~W.~D.; Jain,~A.; Hautier,~G.; Kocher,~M.; Cholia,~S.;
  Gunter,~D.; Chevrier,~V.~L.; Persson,~K.~A.; Ceder,~G. {Python Materials
  Genomics (pymatgen): A robust, open-source python library for materials
  analysis}. \emph{Computational Materials Science} \textbf{2013}, \emph{68},
  314--319\relax
\mciteBstWouldAddEndPuncttrue
\mciteSetBstMidEndSepPunct{\mcitedefaultmidpunct}
{\mcitedefaultendpunct}{\mcitedefaultseppunct}\relax
\EndOfBibitem
\bibitem[Alvarez(2013)]{alvarez}
Alvarez,~S. A cartography of the van der {Waals} territories. \emph{Dalton
  Transactions} \textbf{2013}, \emph{42}, 8617--8636\relax
\mciteBstWouldAddEndPuncttrue
\mciteSetBstMidEndSepPunct{\mcitedefaultmidpunct}
{\mcitedefaultendpunct}{\mcitedefaultseppunct}\relax
\EndOfBibitem
\bibitem[Togo()]{spglib}
Togo,~A. http://spglib.sourceforge.net\relax
\mciteBstWouldAddEndPuncttrue
\mciteSetBstMidEndSepPunct{\mcitedefaultmidpunct}
{\mcitedefaultendpunct}{\mcitedefaultseppunct}\relax
\EndOfBibitem
\bibitem[Hundt \latin{et~al.}(2006)Hundt, Sch{\"o}n, and Jansen]{CMPZ}
Hundt,~R.; Sch{\"o}n,~J.~C.; Jansen,~M. {CMPZ--an algorithm for the efficient
  comparison of periodic structures}. \emph{Journal of applied crystallography}
  \textbf{2006}, \emph{39}, 6--16\relax
\mciteBstWouldAddEndPuncttrue
\mciteSetBstMidEndSepPunct{\mcitedefaultmidpunct}
{\mcitedefaultendpunct}{\mcitedefaultseppunct}\relax
\EndOfBibitem
\bibitem[Lee \latin{et~al.}(2010)Lee, Murray, Kong, Lundqvist, and
  Langreth]{lee2010}
Lee,~K.; Murray,~{\'E}.~D.; Kong,~L.; Lundqvist,~B.~I.; Langreth,~D.~C.
  {Higher-accuracy van der Waals density functional}. \emph{Physical Review B}
  \textbf{2010}, \emph{82}, 081101\relax
\mciteBstWouldAddEndPuncttrue
\mciteSetBstMidEndSepPunct{\mcitedefaultmidpunct}
{\mcitedefaultendpunct}{\mcitedefaultseppunct}\relax
\EndOfBibitem
\bibitem[Cooper(2010)]{cooper2010}
Cooper,~V.~R. {Van der Waals density functional: An appropriate exchange
  functional}. \emph{Physical Review B} \textbf{2010}, \emph{81}, 161104\relax
\mciteBstWouldAddEndPuncttrue
\mciteSetBstMidEndSepPunct{\mcitedefaultmidpunct}
{\mcitedefaultendpunct}{\mcitedefaultseppunct}\relax
\EndOfBibitem
\bibitem[Hamada and Otani(2010)Hamada, and Otani]{hamada2010}
Hamada,~I.; Otani,~M. {Comparative van der Waals density-functional study of
  graphene on metal surfaces}. \emph{Physical Review B} \textbf{2010},
  \emph{82}, 153412\relax
\mciteBstWouldAddEndPuncttrue
\mciteSetBstMidEndSepPunct{\mcitedefaultmidpunct}
{\mcitedefaultendpunct}{\mcitedefaultseppunct}\relax
\EndOfBibitem
\bibitem[Vydrov and Van~Voorhis(2009)Vydrov, and Van~Voorhis]{vydrov2009}
Vydrov,~O.~A.; Van~Voorhis,~T. {Nonlocal van der Waals density functional made
  simple}. \emph{Physical Review Letters} \textbf{2009}, \emph{103},
  063004\relax
\mciteBstWouldAddEndPuncttrue
\mciteSetBstMidEndSepPunct{\mcitedefaultmidpunct}
{\mcitedefaultendpunct}{\mcitedefaultseppunct}\relax
\EndOfBibitem
\bibitem[Vydrov and Van~Voorhis(2010)Vydrov, and Van~Voorhis]{vydrov2010}
Vydrov,~O.~A.; Van~Voorhis,~T. {Nonlocal van der Waals density functional: The
  simpler the better}. \emph{The Journal of chemical physics} \textbf{2010},
  \emph{133}, 244103\relax
\mciteBstWouldAddEndPuncttrue
\mciteSetBstMidEndSepPunct{\mcitedefaultmidpunct}
{\mcitedefaultendpunct}{\mcitedefaultseppunct}\relax
\EndOfBibitem
\bibitem[Sabatini \latin{et~al.}(2013)Sabatini, Gorni, and
  de~Gironcoli]{sabatini2013}
Sabatini,~R.; Gorni,~T.; de~Gironcoli,~S. {Nonlocal van der Waals density
  functional made simple and efficient}. \emph{Physical Review B}
  \textbf{2013}, \emph{87}, 041108\relax
\mciteBstWouldAddEndPuncttrue
\mciteSetBstMidEndSepPunct{\mcitedefaultmidpunct}
{\mcitedefaultendpunct}{\mcitedefaultseppunct}\relax
\EndOfBibitem
\bibitem[Bj{\"o}rkman \latin{et~al.}(2012)Bj{\"o}rkman, Gulans, Krasheninnikov,
  and Nieminen]{bjorkman2012}
Bj{\"o}rkman,~T.; Gulans,~A.; Krasheninnikov,~A.~V.; Nieminen,~R.~M. van der
  {Waals} bonding in layered compounds from advanced density-functional
  first-principles calculations. \emph{Physical Review Letters} \textbf{2012},
  \emph{108}, 235502\relax
\mciteBstWouldAddEndPuncttrue
\mciteSetBstMidEndSepPunct{\mcitedefaultmidpunct}
{\mcitedefaultendpunct}{\mcitedefaultseppunct}\relax
\EndOfBibitem
\bibitem[McGuire(2017)]{McGuire2017}
McGuire,~M.~A. Crystal and Magnetic Structures in Layered, Transition Metal
  Dihalides and Trihalides. \emph{Crystals} \textbf{2017}, \emph{7}\relax
\mciteBstWouldAddEndPuncttrue
\mciteSetBstMidEndSepPunct{\mcitedefaultmidpunct}
{\mcitedefaultendpunct}{\mcitedefaultseppunct}\relax
\EndOfBibitem
\bibitem[Sohier \latin{et~al.}(2015)Sohier, Calandra, and Mauri]{Sohier2015}
Sohier,~T.; Calandra,~M.; Mauri,~F. Density-functional calculation of static
  screening in two-dimensional materials: The long-wavelength dielectric
  function of graphene. \emph{Phys. Rev. B} \textbf{2015}, \emph{91},
  165428\relax
\mciteBstWouldAddEndPuncttrue
\mciteSetBstMidEndSepPunct{\mcitedefaultmidpunct}
{\mcitedefaultendpunct}{\mcitedefaultseppunct}\relax
\EndOfBibitem
\bibitem[Perdew \latin{et~al.}(1996)Perdew, Burke, and Ernzerhof]{PBE}
Perdew,~J.~P.; Burke,~K.; Ernzerhof,~M. Generalized Gradient Approximation Made
  Simple. \emph{Physical Review Letters} \textbf{1996}, \emph{77},
  3865--3868\relax
\mciteBstWouldAddEndPuncttrue
\mciteSetBstMidEndSepPunct{\mcitedefaultmidpunct}
{\mcitedefaultendpunct}{\mcitedefaultseppunct}\relax
\EndOfBibitem
\bibitem[Li \latin{et~al.}(2016)Li, Chen, Shi, and Li]{MYLi2016}
Li,~M.-Y.; Chen,~C.-H.; Shi,~Y.; Li,~L.-J. Heterostructures based on
  two-dimensional layered materials and their potential applications.
  \emph{Materials Today} \textbf{2016}, \emph{19}, 322--335\relax
\mciteBstWouldAddEndPuncttrue
\mciteSetBstMidEndSepPunct{\mcitedefaultmidpunct}
{\mcitedefaultendpunct}{\mcitedefaultseppunct}\relax
\EndOfBibitem
\bibitem[Cheng \latin{et~al.}(2018)Cheng, Guo, Han, Jiang, Zhang, Ahuja, Su,
  and Zhao]{CKai2018}
Cheng,~K.; Guo,~Y.; Han,~N.; Jiang,~X.; Zhang,~J.; Ahuja,~R.; Su,~Y.; Zhao,~J.
  2D lateral heterostructures of group-III monochalcogenide: Potential
  photovoltaic applications. \emph{Applied Physics Letters} \textbf{2018},
  \emph{112}, 143902\relax
\mciteBstWouldAddEndPuncttrue
\mciteSetBstMidEndSepPunct{\mcitedefaultmidpunct}
{\mcitedefaultendpunct}{\mcitedefaultseppunct}\relax
\EndOfBibitem
\bibitem[Castellanos-Gomez \latin{et~al.}(2022)Castellanos-Gomez, Duan, Fei,
  Gutierrez, Huang, Huang, Quereda, Qian, Sutter, and
  Sutter]{Castellanos-Gomez2022}
Castellanos-Gomez,~A.; Duan,~X.; Fei,~Z.; Gutierrez,~H.~R.; Huang,~Y.;
  Huang,~X.; Quereda,~J.; Qian,~Q.; Sutter,~E.; Sutter,~P. Van der Waals
  heterostructures. \emph{Nature Reviews Methods Primers} \textbf{2022},
  \emph{2}, 58\relax
\mciteBstWouldAddEndPuncttrue
\mciteSetBstMidEndSepPunct{\mcitedefaultmidpunct}
{\mcitedefaultendpunct}{\mcitedefaultseppunct}\relax
\EndOfBibitem
\bibitem[van~der Zande \latin{et~al.}(2014)van~der Zande, Kunstmann, Chernikov,
  Chenet, You, Zhang, Huang, Berkelbach, Wang, Zhang, Hybertsen, Muller,
  Reichman, Heinz, and Hone]{vanderZande2014}
van~der Zande,~A.~M.; Kunstmann,~J.; Chernikov,~A.; Chenet,~D.~A.; You,~Y.;
  Zhang,~X.; Huang,~P.~Y.; Berkelbach,~T.~C.; Wang,~L.; Zhang,~F.;
  Hybertsen,~M.~S.; Muller,~D.~A.; Reichman,~D.~R.; Heinz,~T.~F.; Hone,~J.~C.
  Tailoring the Electronic Structure in Bilayer Molybdenum Disulfide via
  Interlayer Twist. \emph{Nano Letters} \textbf{2014}, \emph{14},
  3869--3875\relax
\mciteBstWouldAddEndPuncttrue
\mciteSetBstMidEndSepPunct{\mcitedefaultmidpunct}
{\mcitedefaultendpunct}{\mcitedefaultseppunct}\relax
\EndOfBibitem
\bibitem[Shi \latin{et~al.}(2012)Shi, Zhou, Lu, Fang, Lee, Hsu, Kim, Kim, Yang,
  Li, Idrobo, and Kong]{Shi2012}
Shi,~Y.; Zhou,~W.; Lu,~A.-Y.; Fang,~W.; Lee,~Y.-H.; Hsu,~A.~L.; Kim,~S.~M.;
  Kim,~K.~K.; Yang,~H.~Y.; Li,~L.-J.; Idrobo,~J.-C.; Kong,~J. van der Waals
  Epitaxy of MoS2 Layers Using Graphene As Growth Templates. \emph{Nano
  Letters} \textbf{2012}, \emph{12}, 2784--2791\relax
\mciteBstWouldAddEndPuncttrue
\mciteSetBstMidEndSepPunct{\mcitedefaultmidpunct}
{\mcitedefaultendpunct}{\mcitedefaultseppunct}\relax
\EndOfBibitem
\bibitem[Iannaccone \latin{et~al.}(2018)Iannaccone, Bonaccorso, Colombo, and
  Fiori]{Iannaccone2018}
Iannaccone,~G.; Bonaccorso,~F.; Colombo,~L.; Fiori,~G. Quantum engineering of
  transistors based on 2D materials heterostructures. \emph{Nature
  Nanotechnology} \textbf{2018}, \emph{13}, 183--191\relax
\mciteBstWouldAddEndPuncttrue
\mciteSetBstMidEndSepPunct{\mcitedefaultmidpunct}
{\mcitedefaultendpunct}{\mcitedefaultseppunct}\relax
\EndOfBibitem
\bibitem[Lazic(2015)]{Lazic2015}
Lazic,~P. CellMatch: Combining two unit cells into a common supercell with
  minimal strain. \emph{Computer Physics Communications} \textbf{2015},
  \emph{197}, 324--334\relax
\mciteBstWouldAddEndPuncttrue
\mciteSetBstMidEndSepPunct{\mcitedefaultmidpunct}
{\mcitedefaultendpunct}{\mcitedefaultseppunct}\relax
\EndOfBibitem
\bibitem[Koda \latin{et~al.}(2016)Koda, Bechstedt, Marques, and
  Teles]{Koda2016}
Koda,~D.~S.; Bechstedt,~F.; Marques,~M.; Teles,~L.~K. Coincidence Lattices of
  2D Crystals: Heterostructure Predictions and Applications. \emph{The Journal
  of Physical Chemistry C} \textbf{2016}, \emph{120}, 10895--10908\relax
\mciteBstWouldAddEndPuncttrue
\mciteSetBstMidEndSepPunct{\mcitedefaultmidpunct}
{\mcitedefaultendpunct}{\mcitedefaultseppunct}\relax
\EndOfBibitem
\end{mcitethebibliography}

\end{document}